\newcommand{\eqn}[1]{(\ref{#1})}
\newcommand{\ft}[2]{{\textstyle\frac{#1}{#2}}}
\def\be{\begin{equation}}
\def\ee{\end{equation}}
\def\bea{\begin{eqnarray}}
\def\eea{\end{eqnarray}}
\def\nn{\nonumber}

\def\nn{{\nonumber}}

\renewcommand{\a}{\alpha}
\renewcommand{\b}{\beta}
\def\vep{\varepsilon}                         

\renewcommand{\d}{\delta}
\newcommand{\pa}{\partial}
\newcommand{\g}{\gamma}
\newcommand{\G}{\Gamma}
\newcommand{\e}{\epsilon}

\renewcommand{\l}{\lambda}
\newcommand{\m}{\mu}
\newcommand{\n}{\nu}

\newcommand{\s}{\sigma}

\renewcommand{\o}{\omega}
\newcommand{\q}{\theta}

\newcommand{\del}{\partial}
\def\IZ{{\hbox{{\rm Z}\kern-.4em\hbox{\rm Z}}}}

\def\bigone{{\hbox{1\kern -.23em{\rm l}}}}
\newcommand{\NPB}[3]{{Nucl.\ Phys.} {\bf B#1} (#2) #3}

\newcommand{\PRD}[3]{{ Phys.\ Rev.}\ {\bf D#1} (#2) #3}
\newcommand{\PLB}[3]{{Phys.\ Lett.}\ {\bf B#1} (#2) #3}
\newcommand{\JHEP}[3]{{JHEP} {\bf #1} (#2) #3}
\newcommand{\tmath}[1]{\mbox{$#1$}}

\newcommand{\refer}[1]{(\ref{#1})}
\newcommand{\qb}{\bar{\q}}
\documentstyle[12pt,a4]{article}
\begin{document}
\begin{titlepage}
\begin{flushright} THU-99/05\\ {\tt hep-th/9902051}\end{flushright}
\begin{center}
\vskip 12 mm
{\Large{\bf Supermembranes and Super Matrix Models}}
\vskip 10mm
Bernard de Wit
\vskip 8mm
Institute for Theoretical Physics, Utrecht University\\
Princetonplein 5, 3508 TA Utrecht, The Netherlands
\end{center}
\vskip 12mm
\begin{center}{\bf ABSTRACT} \end{center}
\begin{quotation}\noindent
We review recent developments in the theory of supermembranes and 
their relation to matrix models.
\\
{\small 
\noindent Lecture presented at the Corfu Workshop, September 20 - 26,
1998, of the TMR Project {\it Quantum Aspects of Gauge Theories,
Supersymmetry and Unification} (ERBFMRXCT96-0045), to appear in  the
proceedings.} 
\end{quotation}

\vfill
January 1999
\end{titlepage}
\eject
%
%
%
%
%
%
%
%
\section{Introduction} 
Supermembranes \cite{BST} were proposed as a consistent quantum-mechanical
extension of 11-dimensional supergravity \cite{CJS}, inspired by 
the way in which 
string theory defines a quantum-mechanical extension of
10-dimensional supergravity theories. Although there are 
similarities in the theoretical description of superstrings and
supermembranes, there are also a number of features that are 
distinctly different. An elementary superstring can be formulated as a field 
theory on the 2-dimensional worldsheet swept out by the string in 
Minkowski space. This field theory is free and describes an 
infinite number of 
states with a discrete equidistant mass spectrum with steps
measured by $1/\sqrt{\a^\prime}$, the fundamental mass scale of string
theory. Likewise the supermembrane theory can be formulated as a 
field theory on a 3-dimensional world volume. But unlike the 
previous case, this theory is not a
free but an interacting theory of a complicated structure. 
Furthermore the mass spectrum of the supermembrane is continuous, 
rather than discrete \cite{DWLN}. This is not a generic feature of
quantized extended objects, but crucially rests on the presence of
supersymmetry. At an early stage the question was  
raised whether, in view of Haag's theorem, the supermembrane 
should not be regarded as a second- rather than a first-quantized 
theory, with a unitarily nonimplementable evolution matrix 
\cite{DWHN}.
As it turns out, both issues are resolved in the context of a more
recent perspective  
in which the continuity of the spectrum is seen as arising from 
multi-membrane states. The theory, set up initially to define a 
first-quantized supermembrane, captures also the presence of  
multi-membrane states as described in a second-quantized theory. Again
this feature strongly hinges on supersymmetry: for the generic theory
there is not reason why  states of several {\it interacting} membranes
should give rise to a continuous mass spectrum. 

The continuity of the supermembrane spectrum is due to the fact 
that, quantum-mechanically, the supermembrane can develop stringlike 
zero-area `spikes' which do not contribute to the mass. 
Consequently a membrane can be pinched into two or more 
membranes connected by these stringlike configurations of arbitrary
length, which become indistinguishable from the multi-membrane 
state obtained by suppressing the connecting strings. In this 
way, not only are single- and multi-membrane  
states indistinguishable, but so are certain states of different
topology and states with and without winding (so that topology
changes will correspond to smooth transitions in the moduli space that
parametrizes these states). Thus the first-quantized  
theory of spherical supermembranes ultimately describes also
membranes of nontrivial topology, multi-membrane states and (if the
target space has compact coordinates) supermembranes with winding.

In 11 spacetime dimensions the supermembrane can consistently
couple to a superspace background that satisfies a number of
constraints which are equivalent to the supergravity equations
of motion. The supermembrane action
can also exist in $4,5$ and 7 spacetime dimensions, 
in the same way as the Green-Schwarz superstring
\cite{GS84} is classically consistent in $3,4,6$ and 10
dimensions.  In the context of string theory it was natural to 
expect that the massless states of the supermembrane would 
correspond to those of 11-dimensional 
supergravity. However, in the presence of a continuous mass
spectrum \cite{DWLN} the  possible existence of massless
states is difficult to prove or disprove
\cite{DWHN,DWN,FH}. The unability to make sense of the mass spectrum
and the fact that 11-dimensional supergravity seemed to have no place
in string theory, formed an obstacle for further development of the
theory. More recently, however, interest in supermembranes was 
rekindled by the realization that 11-dimen\-sion\-al supergravity 
does have its role to play as the long-distance approximation to M-theory
\cite{HT,Townsend,witten3,horvw}.  M-theory is the
conjectured framework for unifying  
all five superstring theories and 11-dimensional supergravity. It
turns out that supermembranes, M-theory and super-matrix-models are all
intricately related. 

An important observation was that it is possible to regularize 
the supermembrane in terms of a super matrix model based on some 
finite group, such as U($N$). In the limit of infinite $N$ one 
then recovers the supermembrane \cite{DWHN}. These supersymmetric
matrix models were  
constructed long ago \cite{CH} and can be obtained from 
supersymmetric Yang-Mills theories in the zero-volume 
limit. More recently it was realized that these models describe 
the short-distance behaviour of $N$ Dirichlet particles 
\cite{boundst}. The 
continuity of the spectrum is then understood directly in terms of 
the spectrum of $N$-particle states. A bold conjecture was that these
super matrix models capture the degrees of freedom of M-theory 
\cite{BFSS}. In the large-$N$ limit, where one considers the 
states with an infinite number of particles, the supermembranes 
should then re-emerge. Furthermore 
there is evidence meanwhile that the supermembrane has massless states
\cite{mboundstates},
which will presumably correspond to the states of 11-dimensional
supergravity, although proper asymptotic states do not exist. The
evidence is based on the matrix model regularization of the
supermembrane for low values of $N$. For fixed value of $N$ the 
existence of such states was foreseen on the basis of identifying
the 
Kaluza-Klein states of M-theory compactified on $S^1$ with the
Dirichlet particles and their bound states in type-IIA string
theory. 

{From} this viewpoint it is natural to consider the
supermembrane in curved backgrounds
associated with 11-dimensional supergravity. Such backgrounds
consist of a nontrivial metric, a three-index gauge
field and a gravitino field. This provides us with an
action that transforms as a scalar under the combined (local) 
supersymmetry transformations of the background fields and the
supermembrane embedding coordinates. Here it is important to
realize that the supersymmetry transformations of the embedding
coordinates will themselves depend on the background. When
the background is supersymmetric, then the action will be
supersymmetric as well. 
In the light-cone formulation this model will lead to models
invariant under area-preserving diffeomorphisms, which in
certain situations can be approximated by matrix models 
in curved backgrounds. The area-preserving diffeomorphisms
are then replaced by a finite group, such as U$(N)$, but 
target-space diffeomorphisms are no longer manifestly
realized. Matrix 
models in curved space have already been studied in 
\cite{oldBG}. Recently toroidal compactifications of matrix
theory were considered in which the three-form gauge field of
11-dimensional gravity plays a crucial role \cite{newBG}. These
compactifications exhibit interesting features in which the
noncommutative torus appears as a new solution to compactified matrix
theory.  We should also point out that 
classical supermembrane solutions in nontrivial 
backgrounds have been discussed before, see,
e.g. \cite{BDPS}. In view of the relation between near-horizon 
geometries and conformal field theories \cite{maldacena} 
interesting classes of backgrounds are the ones where the target 
space factorizes locally into the product of an $AdS$ space and some
compact space.  

In this lecture we review many of these topics 
starting from the supermembrane point of view. We should stress 
that there remain many open questions and problems, both for 
supermembranes and for super matrix models. For 
instance, the large-$N$ behaviour is still poorly understood as 
are features related to matrix models and membranes in nontrivial 
backgrounds. But the most intriguing questions concern the precise
role that these theories play in M-theory, the theory that encompasses
all known perturbative string theories. For other reviews, we refer to
\cite{review}; a number of related topics was also discussed in the
workshop and can be found in this volume. 

\section{Supermembranes}                         
Fundamental supermembranes can be described in terms of actions 
of the Green-Schwarz type, possibly in a nontrivial but 
restricted (super)spacetime background \cite{BST}. Such actions exist 
for supersymmetric $p$-branes, where $p= 0, 1,\ldots, d-1$ defines the
spatial dimension of the brane. Thus for
$p=0$ we have a superparticle, for $p=1$ a superstring, for
$p=2$ a supermembrane, and so on. The
dimension of spacetime in which the superbrane can live is
very restricted. These restrictions arise from the 
fact that the action contains a 
Wess-Zumino-Witten term, whose supersymmetry depends sensitively
on the spacetime dimension. If the coefficient of this term
takes a particular value then the action possesses an additional
fermionic gauge symmetry, the so-called $\kappa$-symmetry.
This symmetry is necessary to
ensure the matching of (physical) bosonic and fermionic
degrees of freedom.
In the following we restrict ourselves to 
supermembranes (i.e., $p=2$) in 11 dimensions.  

The 11-dimensional supermembrane \cite{BST} is written in terms of 
superspace embedding coordinates 
$Z^M(\zeta)=(X^\mu(\zeta),\theta^\alpha(\zeta))$, which are functions of 
the three world-volume coordinates $\zeta^i$ ($i = 0,1,2$). It 
couples to the superspace geometry of 11-dimensional supergravity,
encoded by the supervielbein $E_M{}^A$ and the antisymmetric tensor
gauge superfield $B_{MNP}$, through the action\footnote{%
  Our notation and conventions are as follows. Tangent-space indices are
  $A=(r,a)$, whereas curved indices are denoted by
  $M=(\mu,\alpha)$. Here $r,\mu$ refer to  commuting and
  $a,\alpha$ to anticommuting coordinates. Moreover we take 
  $\epsilon_{012}=-\epsilon^{012}=1$. } %
\be
S[Z(\zeta)] = \int {\rm d}^3\zeta \;\Big[- \sqrt{-g(Z(\zeta))} - \ft16
\varepsilon^{ijk}\, \Pi_i^A \Pi_j^B\Pi_k^C \,B_{CBA}(Z(\zeta)) 
\,\Big]\,,
\label{supermem}
\ee
where $\Pi^A_i = \pa Z^M/\pa\zeta^i \; E_M{}^{\!A}$ is the pull-back of
the supervielbein to the membrane worldvolume. Here the 
induced metric equals $g_{ij} =\Pi^r_i \Pi^s_j \,\eta_{rs}$, with
$\eta_{rs}$ being the constant Lorentz-invariant metric. This action
is invariant under local fermionic $\kappa$ transformations 
alluded to above,
given that certain constraints on the background fields hold, which are 
equivalent to the  equations of motion of 11-dimensional 
supergravity \cite{BST}.

Flat superspace is characterized by
\be\label{flatssquantities}
\begin{array}{rclcrcl}
E_\mu{}^r &=& \d_\mu{}^r \, , & & E_\mu{}^a &=&0  \, , \\
E_\alpha{}^a &=& \delta_\alpha{}^a  \, ,& & 
E_\alpha{}^r &=& -(\bar\theta \Gamma^r)_{\alpha} \, , \\
B_{\mu\n\alpha} &=& (\bar\theta\Gamma_{\m\n})_{\alpha}\,
, &&
B_{\m\a\b} &=& 
(\bar\theta\Gamma_{\m\n})_{(\a}\,(\bar\theta\Gamma^\n)_{\b)} \,, \\
 B_{\a\b\g} &=& (\bar\theta\Gamma_{\m\n})_{(\a}\, 
(\bar\theta\Gamma^\m)_{\b}\, 
(\bar\theta\Gamma^\n)_{\g)}\,,&\hspace{6.2mm}  &
B_{\mu\nu\rho} &=&0  \, .  
\end{array}
\ee
The gamma matrices 
are denoted by $\Gamma^r$; gamma matrices 
with more than one index denote antisymmetrized products of gamma
matrices with unit weight. In flat  
superspace the supermembrane Lagrangian, written in 
components, reads (in the notation and conventions of \cite{DWHN}),  
\bea
&&{\cal L} =- \,\sqrt{-g(X,\theta)} - \varepsilon^{ijk} \,
\bar\theta\Gamma_{\mu\nu}\partial_k\theta \Big [{\textstyle{1\over 2}}
\,\partial_i X^\mu (\partial_j X^\nu +
\bar\theta\Gamma^\nu\partial_j \theta) + {\textstyle{1\over 6}}
\,\bar\theta\Gamma^\mu\partial_i\theta\;
\bar\theta\Gamma^\nu\partial_j\theta \Big] \,, \qquad{}
\label{action}
\eea
The target space can 
have compact dimensions which permit winding membrane states 
\cite{DWPP1}. In flat superspace the induced metric,
\be
g_{ij} = (\partial_iX^\mu + \bar\theta\Gamma^\mu
\partial_i\theta) (\partial_jX^\nu +
\bar\theta\Gamma^\nu
\partial_j\theta) \,\eta_{\mu\nu}\,,
\ee
is supersymmetric. Therefore the first term in 
\eqn{action} is trivially invariant under spacetime
supersymmetry,
\be
\d X^\mu = -\bar\e \G^\m\theta \,, \qquad \d\theta = \e\,.
\ee 
In $4$, $5$, $7$, or $11$ spacetime dimensions the 
second term in the action proportional to $\varepsilon^{ijk}$ is also 
supersymmetric (up to a total divergence) and the full action is 
invariant under  $\kappa$-symmetry.
 
In the case of the open supermembrane, $\kappa$-symmetry imposes 
boundary conditions on the fields \cite{open1,open2}. They must 
ensure that the following integral over the  
boundary of the membrane world volume vanishes,
\bea
&&\int_{\del M} \Big[ \ft12{\rm d} X^\m \wedge ( {\rm d}X^\n + 
\bar\theta\G^\nu {\rm d}\theta)\, \bar \theta \G_{\m\n} \d_\kappa 
\theta  +\ft16\bar\theta\G^\mu {\rm d}\theta\wedge \bar\theta\G^\nu {\rm 
d}\theta\, \bar\theta\G_{\m\nu} \d_{\kappa}\theta \nonumber\\
&&\hspace{40mm} +\ft12( {\rm d}X^\m - \ft13 \bar\theta\G^\mu {\rm 
d}\theta ) \wedge 
\bar\theta\G_{\m\nu} {\rm d}\theta\; \bar\theta\G^\n \d_{\kappa}\theta
\Big ]=0 \,. \label{kappaboundary}
\eea
This can be achieved by having a ``membrane D-$p$-brane'' at 
the boundary with $p=1,5$, or 9, which is defined in terms of 
$(p+1)$ Neumann and $(10-p)$ Dirichlet boundary 
conditions for the $X^\mu$, together with corresponding boundary 
conditions on the fermionic coordinates.  
More explicitly, we define projection operators
\be 
{\cal P}_\pm=\ft12\Big({\bf 1} \pm \G^{p+1}\, \G^{p+2}\cdots 
\G^{10}\Big)\,, \label{projectors}
\ee
and impose the Dirichlet boundary conditions
\bea
\del_\parallel \, X^M\big|&=& 0\,, \qquad M=p+1,\ldots,10\,, \nonumber\\
{\cal P}_- \q\big|&=&0\, , \label{boundcond}
\eea 
where $\del_\perp$ and $\del_\parallel$ define the world-volume 
derivatives perpendicular or tangential to the surface swept out 
by the membrane boundary in the target space. Note that the fermionic 
boundary condition implies that ${\cal P}_- \del_\parallel\q=0$. 
Furthermore, it implies that spacetime 
supersymmetry is reduced to only 16 supercharges associated with 
spinor parameters ${\cal P}_+\epsilon$, which is {\it chiral} with 
respect to the ($p+1$)-dimensional world volume of the 
D-$p$-brane at the boundary. With respect to 
this reduced supersymmetry, the superspace coordinates decompose 
into two parts, one corresponding to $(X^M, {\cal P}_-\theta)$ and the 
other corresponding to $(X^m, {\cal P}_+\theta)$ where
$m=0,1,\ldots,p$. While for the 
five-brane these superspaces exhibit a somewhat balanced decomposition in 
terms of an equal number of bosonic and fermionic coordinates,
the situation for $p=1,9$ shows heterotic features in that 
one space has an excess of fermionic and the other an excess of 
bosonic coordinates. Moreover, we note that supersymmetry may be further
broken, e.g.\ by choosing different Dirichlet conditions 
on nonconnected segments of the supermembrane boundary.

The Dirichlet boundary conditions can be supplemented by the 
following Neumann boundary conditions,
\bea
\del_\perp \, X^m\big|&=& 0 \qquad m=0,1,\ldots,p \,,\nonumber\\
{\cal P}_+ \del_\perp \q \big|&=&0 \,. \label{Nboundcond}
\eea 
These do not lead to a further breakdown of the rigid spacetime 
symmetries.

We now continue and follow the light-cone quantization described 
in \cite{DWHN} for a closed membrane without winding. In the
light-cone gauge  
the light-cone coordinate $X^+=(X^{1}+X^0)/\sqrt2$ is linearly
identified with to the world-volume time denoted by $\tau$ and the
fermionic coordinates are subject to the gauge  
condition \tmath{\g_+\, \q=0}. The momentum $P_-$ is time 
independent and proportional to the center-of-mass (CM) value $P^+_0= 
(P_-)_0$ times some  
density ${\sqrt{w(\s)}}$ of the spacesheet, whose spacesheet 
integral is normalized to unity. Hence we have
\be
P^+_0 = \int {\rm d}^2\! \sigma \,P^+(\s) . 
\ee
The CM momentum
$P_0^-$ is equal to minus the Hamiltonian and takes the form 
\begin{eqnarray}
\label{memham}
H&=&  \frac{1}{P_0^+}\, \int {\rm d}^2\s \, \sqrt{w}\,
\bigg[ \, \frac{P^a\, P_a }{2\,w} + \ft{1}{4} \{\, 
X^a,X^b\,\}^2 -P^+_0\, \qb\,\g_- 
\g_a\, \{\, X^a , \q\,\}\, \bigg]\, .
\end{eqnarray}
Here the integral runs over the spatial components of the
world volume denoted by $\s^1$ and $\s^2$, while 
$P^a(\s)$ ($a=2,\ldots,9$) are the momenta conjugate to the 
transverse coordinates $X^a$. Furthermore we made use of the
Poisson bracket \tmath{\{ A,B\} } defined by
\be
\{ A(\s ),B(\s )\} = \frac{1}{\sqrt{w(\s)}}\, \varepsilon^{rs}\, 
\del_r A(\s )\, 
\del_s B(\s ). \label{poisbrak}
\ee
Note that the coordinate $X^-=(X^{1}-X^0)/\sqrt2$ itself does
not appear in the Hamiltonian \refer{memham}. It is defined via 
\be
P^+_0\, \del_rX^-= - \frac{{\bf P} \cdot \del_r{\bf X}}{\sqrt{w}} - 
P^+_0\, \qb\g_-\del_r\q\,, \label{delxminus}
\ee
and implies that the right-hand side of \eqn{delxminus} 
must be closed; without winding in $X^-$, it must be 
exact. This constraint is important later on. 

The other CM coordinates and momenta are    
\be
{\bf P}_0 = \int {\rm d}^2\! \sigma \; {\bf P} \,,  \qquad {\bf 
X}_0 = \int \!{\rm d}^2\!\sigma\sqrt{w(\sigma)}\, {\bf X}(\sigma)\,, 
\qquad 
\theta_0 = \int \!{\rm d}^2\!\sigma 
\sqrt{w(\sigma)}\, \theta(\sigma)\,.  
\ee
In the light-cone gauge we are left with the transverse
coordinates $\bf X$ and corresponding momenta $\bf P$, which
transform as vectors under the SO(9) group of transverse
rotations. Only
sixteen fermionic components $\theta$ remain, which transform as
SO(9) spinors. Furthermore we have the CM momentum
$P_0^+$ and the center-of-mass coordinate $X^-_0$ (the
remaining modes in $X^-$ are dependent).

The supermembrane Hamiltonian \eqn{memham} can be decomposed as
follows, 
\be
H = {{\bf P}_0^{\,2}\over 2 P_0^+} + {{\cal M}^2\over 2P_0^+}  
\,. \label{mbhamiltonian}
\ee 
Because the Hamiltonian is equal to $- P_0^-$, $\cal M$ is the
supermembrane mass operator, which does {\it 
not} depend on any of the CM coordinates or momenta. The explicit 
expression for ${\cal M}^2$ is 
\be
{\cal M}^2 =  \int {\rm d}^2\!\s \; \sqrt{w(\s)} \bigg[ {[{\bf 
P}^2(\s)]' \over w(\s)}  +\ft12 \Big(\{X^a,X^b\}\Big)^2 - 2 P_0^+ 
\, \bar\theta\g_- \g_a\{X^a,\theta\}\bigg]\,, \label{mass}
\ee
where $[{\bf P}^2]^\prime$ indicates that the contribution of the CM 
momentum ${\bf P}_0$ is suppressed. 

The structure of the Hamiltonian \eqn{mbhamiltonian} shows that the wave
functions for the supermembrane now factorize into a
wave function of the CM modes and a wave
function of the supersymmetric quantum-mechanical system that
describes the other modes. For the latter the mass operator plays 
the role of the Hamiltonian of a supersymmetric model in quantum
mechanics. The aspects related to supersymmetry will be discussed in
the next section.

In the light-cone gauge there is still a residual invariance
associated with area-preserving diffeomorphisms of 
the membrane spacesheet. These are defined by transformations 
\begin{equation}
\s^r \to \s^r + \xi^r(\s) \,, 
\end{equation}
with
\begin{equation}\label{APD}
\del_r(\sqrt{w(\s)}\, \xi^r(\s)\, )=0.
\end{equation}
It is convenient to rewrite this condition in terms of dual spacesheet 
vectors by  
\be                                  
\sqrt{w(\s)}\,\xi^r(\s)= \varepsilon^{rs}\, \xi_s(\s)\, .\label{1form}
\ee
In the language of differential forms the
condition \refer{APD} then implies that $\xi_r$ corresponds to a
closed one-form. The trivial solutions are the exact forms, in components,  
\be
\xi_r=\del_r\xi(\s)\,,\label{exact}
\ee
for any globally defined function $\xi(\s)$. The nontrivial solutions are
the closed forms which are not exact. On a Riemann surface of
genus $g$ there are precisely $2g$ linearly independent non-exact 
closed forms, whose integrals along the homology cycles are 
normalized to unity. In components we write
\be
\xi_r=\phi_{(\l)\, r}\;, \qquad \l=1,\ldots,2g\,.
\ee

The presence of the closed but non-exact forms is crucial for
describing 
the winding of the embedding coordinates. More precisely, while 
the momenta ${\bf P}(\s)$ and the fermionic coordinates 
$\theta(\s)$ remain single valued on the spacesheet, the 
embedding coordinates, written as one-forms with components 
$\del_r {\bf X}(\s)$ and  $\del_r X^-(\s)$, are decomposed into 
closed one-forms. Their non-exact contributions are multiplied by an 
integer times the length of the compact direction.

\section{Gauge theory of area-preserving diffeomorphisms}
It turns out that the light-cone formulation of the supermembrane 
can be described as a gauge theory of area-preserving 
diffeomorphisms. Under these diffeomorphisms the fields $X^a$,
$X^-$ and $\q$ transform according to
\be
\label{APDtrafoXtheta}
\d X^a= \displaystyle{\varepsilon^{rs}\over \sqrt{w}}\, \xi_r\, \del_s X^a\,,
\quad 
\d X^-= \displaystyle {\varepsilon^{rs}\over \sqrt{w}}\, \xi_r\, \del_s 
X^-\,, 
\quad
\d \q^a= \displaystyle {\varepsilon^{rs}\over \sqrt{w}}\, 
\xi_r\, \del_s \q\,, 
\ee
where the time-dependent reparametrization $\xi_r$ consists of
closed exact and non-exact parts. The commutator of two infinitesimal
area-preserving  diffeomorphisms is determined by the product rule
\begin{equation}
\xi_r^{(3)} = \partial_r \left( \frac{\epsilon^{st}}{\sqrt{w}} 
\xi_s^{(2)}\xi_t^{(1)}\right) \,,
\end{equation}
where both $\xi_r^{(1,2)}$ are closed vectors. Because 
$\xi_r^{(3)}$ is exact, the exact vectors thus 
generate an invariant subgroup of the area-preserving 
diffeomorphisms. As we shall discuss in the next section this 
subgroup can be approximated by SU$(N)$ in the large-$N$ limit, 
at least for closed membranes. For open membranes the boundary conditions 
on the fields \refer{boundcond} lead to a smaller group, such as SO($N$).
Accordingly there is a gauge
field $\o_r$, which is therefore closed as well and transforming 
as 
\be\label{APDtrafoomega}
\d\o_r=\del_0\xi_r + \del_r \bigg( {\varepsilon^{st}\over\sqrt{w}}\,
\xi_s\,\o_t\bigg)\,.
\ee
Corresponding covariant derivatives are 
\be
\label{covderiv}
D_0 X^a = \displaystyle\del_0X^a - {\varepsilon^{rs}\over \sqrt{w}}\, 
\o_r\, \del_s X^a\,, \qquad  
D_0 \q  = \displaystyle \del_0\q - {\varepsilon^{rs}\over \sqrt{w}}\, 
\o_r\, \pa_s\q\,, 
\ee 
and likewise for \tmath{D_0 X^-}. 

The action corresponding to the following Lagrangian density is
then gauge invariant under the 
transformations \refer{APDtrafoXtheta} and \refer{APDtrafoomega},
\bea
\label{gtlagrangian}
{\cal L}&=&P^+_0\,\sqrt{w}\, \Big[\,  
\ft{1}{2}\,(D_0{\bf X})^2 + \qb\,\g_-\,
D_0\q - \ft{1}{4}\,(P^+_0)^{-2}\,  \{ X^a,X^b\}^2 \\
&& \hspace{16mm}  + (P^+_0)^{-1}\, \qb\,\g_-\,\g_a\,\{X^a,\q\} +
  D_0 X^-\Big]\, ,\nonumber 
\eea
where we draw attention to the last term proportional to
$X^-$, which can be dropped in the absence of winding. 
Moreover, we note that for open
supermembranes, \refer{gtlagrangian} is invariant under the 
transformations \refer{APDtrafoXtheta} and \refer{APDtrafoomega} 
only if $\xi_\parallel=0$ holds on the boundary.
This condition defines a subgroup of the 
group of area-preserving transformations, which is consistent 
with the Dirichlet conditions \refer{boundcond}. Observe that 
here $\del_\parallel$ and $\del_\perp$ refer to the {\it 
spacesheet} derivatives tangential and perpendicular to the 
membrane boundary\footnote{%
  Consistency of the Neumann boundary conditions 
  \refer{Nboundcond} with the area-preserving diffeomorphisms 
  \refer{APDtrafoXtheta} further imposes 
  $\partial_\perp\xi^\parallel=0$ 
  on the boundary, where indices are raised according to 
  \refer{1form}.}. %

The action corresponding to
\refer{gtlagrangian} is also invariant under the 
supersymmetry transformations  
\bea
\d X^a &=& -2\, \bar{\e}\, \g^a\, \q\,, \nonumber\\
\d \q  &=& \ft{1}{2} \g_+\, (D_0 X^a\, \g_a + \g_- )\, \e 
+\ft{1}{4}(P^+_0)^{-1} \, \{ X^a,X^b \}\, \g_+\, \g_{ab}\, \e ,
\nonumber\\ 
\d \o_r &=& -2\,(P^+_0)^{-1}\, \bar{\e}\,\pa_r\q\,.
\label{susytrafos}
\eea
The supersymmetry variation of $X^-$ is not relevant and may be
set to zero. For open membranes one finds that the boundary conditions
$\omega_\parallel=0$
and \mbox{$\epsilon={\cal P}_+\,\epsilon$} must be fulfilled
in order for \refer{susytrafos} to be a symmetry of the action.
In that case the theory takes the form of a gauge theory coupled 
to matter. The pure gauge theory is associated with the Dirichlet 
and the matter with the Neumann (bosonic and 
fermionic) coordinates.  

In the case of a `membrane D-$9$-brane' one now sees that the
degrees of freedom  on the `end-of-the world' $9$-brane precisely
match those of 10-dimensional heterotic strings. {\it On} the boundary
we are left with eight propagating bosons $X^m$ (with $m=2,
\ldots,9$), as $X^{10}$ is constant on the boundary 
due to \refer{boundcond},
paired with the 8-dimensional chiral spinors $\theta$ (subject 
to $\g_+ \theta= {\cal P}_-\theta=0$),
i.e., the scenario of Ho\u{r}ava-Witten \cite{horvw}.

The full equivalence with the membrane Hamiltonian is now established by
choosing the $\o_r=0$ gauge and passing to the Hamiltonian 
formalism. The field equations for $\o_r$ then lead to
the membrane constraint \refer{delxminus} (up to exact contributions), 
partially defining \tmath{X^-}.
Moreover the Hamiltonian corresponding to the gauge theory Lagrangian of 
\refer{gtlagrangian} is nothing
but the light-cone supermembrane Hamiltonian \refer{memham}.
Observe that in the above gauge theoretical construction the space-sheet
metric $w_{rs}$ enters only through its density $\sqrt{w}$ and hence
vanishing or singular metric components do not pose problems.

We are now in a position to study the full 11-dimensional supersymmetry
algebra of the winding supermembrane. For this we decompose the
supersymmetry charge $Q$ associated with the transformations 
\refer{susytrafos}, into two 16-component spinors,
\be
Q= Q^+ + Q^- , \quad \mbox{where}\quad
 Q^\pm = \ft{1}{2}\, \g_\pm\,\g_\mp\, Q\,, \label{Qdecomposition}
\ee
to obtain
\bea
Q^+&=&\int {\rm d}^2 \s \, \Big(\, 2\, P^a\, \g_a + \sqrt{w}\, \{\,
X^a, X^b\, \} \, \g_{ab}\, \Big) \, \q \,, \nonumber \\
Q^-&=& 2\, P^+_0\, \int {\rm d}^2\s\, \sqrt{w}\, \g_-\, \q .
\label{Q-cont}
\eea
In the presence of winding the supersymmetry algebra takes the 
form \cite{DWPP1}
\begin{eqnarray}
\label{contsusy}
(\, Q^+_\a, \bar{Q}^+_\b\, )_{\mbox{\tiny DB}} &=& 2\, 
(\g_+)_{\a\b}\, H 
 - 2\,  (\g_a\, \g_+)_{\a\b}\, \int{\rm d}^2\s\, \sqrt{w}\, \{\, 
X^a, X^-\,\}\, , \nonumber \\ 
(\, Q^+_\a, \bar{Q}^-_\b\, )_{\mbox{\tiny DB}} &=& -(\g_a\,\g_+\,
\g_- )_{\a\b}\, P^a_0  
  - \ft{1}{2}\,(\g_{ab}\, \g_+\g_- )_{\a\b}\, \int {\rm d}^2\s\,
\sqrt{w}\, \{\, X^a,X^b\,\}\,,\nonumber\\[1mm] 
(\, Q^-_\a, \bar{Q}^-_\b\, )_{\mbox{\tiny DB}} &=& -2\, (\g_- )_{\a\b}\, P^+_0\, , 
\end{eqnarray}
where use has been made of the Dirac brackets of the phase-space 
variables and the defining equation \refer{delxminus} for $\pa_r X^-$.

The new feature of this supersymmetry algebra is the emergence of the 
central charges in the first two anticommutators, which are
generated through the winding contributions.
They represent topological quantities obtained by integrating
the winding densities 
\begin{equation}
z^{a}(\s)=\varepsilon^{rs}\,\del_r X^a\,\del_s X^-
\end{equation}
and
\begin{equation}
z^{ab}(\s) =\varepsilon^{rs}\,\del_r X^a\,\del_s X^b
\end{equation}
over the space-sheet. It is gratifying to observe the manifest
Lorentz invariance of \refer{contsusy}. Here  we should point out
that, in adopting the light-cone gauge, we assumed that there was
no winding for 
the coordinate $X^+$. In \cite{BSS} the corresponding algebra for
the matrix regularization was studied. 
The result coincides with ours in the
large-$N$ limit, in which an additional longitudinal five-brane
charge vanishes, provided that one identifies the longitudinal
two-brane charge with the central charge in the
first line of \refer{contsusy}. This identification requires the definition of
$X^-$ in the matrix regularization, a topic that we return to in 
the next section. The form of the algebra is another indication of the 
consistency of the supermembrane-supergravity system.

Until now we discussed the general case of a flat target space 
with possible winding states. To make the identification with the 
matrix models more explicit, let us again ignore the winding and 
split off the center-of-mass (CM) variables as in the previous
section. As discussed there the structure of the Hamiltonian
\eqn{mbhamiltonian} shows that the wave functions for the
supermembrane now factorize into a trivial wave function pertaining to
the CM modes and a wave 
function of the supersymmetric quantum-mechanical system that
describes the other (interacting) modes. For the latter the mass
operator plays  
the role of the Hamiltonian. When the mass operator vanishes on the state, 
then the 32 supercharges act exclusively on the CM coordinates 
and generate a massless supermultiplet of eleven-dimensional 
supersymmetry. In case there is no other degeneracy beyond that 
caused by supersymmetry, the resulting supermultiplet is the one 
of supergravity, describing the graviton, the antisymmetric tensor 
and the gravitino. In terms of the $SO(9)$ helicity 
representations, it  
consists of ${\bf 44} \oplus {\bf 84}$ bosonic and $\bf 128$
fermionic states. For an explicit construction of these states, 
see \cite{PW}. 
When the mass operator does not vanish on the states, we are 
dealing with huge supermultiplets consisting of multiples of 
$2^{15}+2^{15}$ states.

\section{The matrix approximation}
One may expand the supermembrane coordinates and momenta
on the spacesheet in a complete set of functions $Y_A$ with 
$A= 0,1,2, \ldots, \infty$. It is convenient to choose $Y_0=1$. 
Furthermore we choose a basis of the closed one-forms, consisting 
of the exact ones, $\pa_r Y_A$, and a set of closed nonexact 
forms denoted by $\phi_{(\l)r}$. 
Completeness of the $Y_A$ implies the following decompositions,
\bea
\{ Y_A, Y_B\} &=& f_{AB}{}^{\!C}\, Y_C\,, \nonumber  \\[1.9mm]
{\varepsilon^{rs}\over \sqrt w} \,\phi_{(\l)r}\,\pa_s Y_A &=& 
f_{\l A}{}^{\!B} \, Y_B\,, \nonumber\\
{\varepsilon^{rs}\over \sqrt w} \,\phi_{(\l)r}\,
\phi_{(\l^\prime)s} &=& f_{\l \l^\prime} {}^{\!A} \, Y_A\,,
\eea
so that the constants $f^{AB}{}_{\!C}$, $f_{\l A}{}^{\!B}$ and 
$f_{\l \l^\prime} {}^{\!A}$ represent the
structure constants of the infinite-dimen\-sional group of 
area-preserving diffeomorphisms. 
Lowering of indices can be done with the help of the 
invariant metric
\begin{equation}
\eta_{AB}= \int {\rm d}^2\s\, \sqrt{w(\s)}\; Y_A(\s)\, Y_B(\s)\,.
\end{equation}
There is no need to introduce a metric for the $\l$ indices. 
Observe that we have $\eta_{00}=1$. Furthermore it is convenient 
to choose the functions $Y_A$ with $A\geq 1$ such that 
$\eta_{0A}=0$. Completeness implies
\be
\eta^{AB}\,Y_A(\s)\,Y_B(\rho) = {1\over \sqrt{w(\s)}}\,
\d^{(2)}(\s,\rho)\,.
\ee
After lowering of upper indices, the structure constants are defined as 
follows \cite{DWMN,DWPP1}, 
\bea
f_{ABC} &=& \int {\rm d}^2\s\, \varepsilon^{rs}\,\del_r Y_A(\s)\, 
\del_s Y_B(\s)\, Y_C(\s)\,, \nonumber \\ 
f_{\l BC} &=& \int {\rm d}^2\s\, \varepsilon^{rs}\,\phi_{(\l)\, 
r}(\s)\, \del_s Y_B(\s)\, Y_C(\s)\,, \nonumber \\ 
f_{\l \l^\prime C} &=& \int {\rm d}^2\s\, \varepsilon^{rs}\,
\phi_{(\l)\, r}(\s)\, \phi_{(\l^\prime)\, s}(\s)\, Y_C(\s) \, . 
\eea 
Note that we have $f_{AB0}=f_{\l B0}=0$.

Using the above basis one may write down the following
mode expansions for the phase-space variables of the
supermembrane, 
\bea
\del_r{\bf X}(\s) &=& \sum_{\l} \,{\bf X}^\l\, \phi_{(\l)\, r}(\s) 
+ \sum_A\, {\bf X}^A\, \del_r Y_A(\s)\,,\nonumber \\
{\bf P}(\s) &=& \sum_A \,\sqrt{w(\s)}\; {\bf P}^A\, Y_A(\s)\,, \nonumber\\
\q(\s) &=& \sum_A \,\q^A\, Y_A(\s) \,, \label{modeexp}
\eea
introducing winding modes for the transverse coordinates $\bf X$. 
A similar expansion exists for $X^-$.  

Other tensors are needed, for instance, to write down the Lorentz 
algebra generators \cite{DWMN}. An obvious tensor is 
given by 
\be
d_{ABC} =  \int {\rm d}^2\s\, \sqrt{w(\s)}\; Y_A(\s)\, Y_B(\s)\, 
Y_C(\s) \,, 
\ee
which is symmetric in all three indices and satisfies $d_{AB0}= 
\eta_{AB}$. 
Another tensor, whose definition is more subtle, arises when 
expressing $X^-$ in terms of the other coordinates and momenta. 
We recall that $X^-$ is restricted by \eqn{delxminus}, which 
implies the following Gauss-type constraint, 
\be
\varphi^A= f_{BC}{}^{\!A}\Big[ {\bf P}^B\cdot {\bf X}^C + P_0^+\, \bar 
\theta^B\g_- \theta^C\Big] + f_{B\l}{}^{\!A} \, {\bf P}^B\cdot {\bf 
X}^\l \approx 0\,.  \label{constraint}
\ee
The coordinate $X^-$ receives contributions proportional to $Y_A(\s)$, 
which can be parametrized by ($A\not=0$)
\be
X^-_A\approx  {1\over 2P_0^+}\, c^A{}_{\!BC}\Big[{\bf P}^B\cdot{\bf X}^C + 
P_0^+\, \bar  \theta^B\g_- \theta^C\Big]  + {1\over 2P_0^+}\, 
c^A{}_{\!B\l}\, {\bf P}^B\cdot{\bf X}^\l\,. \label{X-A}
\ee
In addition $X^-$ has CM and winding modes. Observe that the 
tensors $c^A{}_{\!BC}$ and $c^A{}_{\!B\l}$
are ambiguous, as \eqn{X-A} is only defined up to the 
constraints \eqn{constraint}. The symmetric component of 
$c^A{}_{\!BC}$ is, however, fixed and given by  
$c^A{}_{\!BC}+c^A{}_{\!CB}= 
-2 d_{ABC}$. Note that $c^A{}_{\!B0}=0$. There are many other 
identities between the various tensors which can be derived by using
completeness. Some examples are \cite{DWMN}, 
\bea
&&f_{[AB}{}^{\!E}\, f_{C]E}{}^{\!D} =d_{(AB}{}^{\!E}\, 
f_{C)E}{}^{\!D} = d_{ABC} 
\, f_{[DE}{}^{\!B}\, f_{FG]}{}^{\!C}= \nonumber\\
&&c_{DE}{}^{\![A}f^{BC]E}= 
d_{EA[B}d_{C]D}{}^{\!E} = 0\,. \label{identities}
\eea
The first identity is just the Jacobi identity for the structure 
constants of the group of area-preserving diffeomorphisms and the second
expresses the fact that $d_{ABC}$ is a group-invariant
tensor. 

It is possible to replace the group of the area-preserving diffeomorphisms   
by a {\it finite} group, so that \eqn{mass} defines the 
Hamiltonian of a supersymmetric quantum-mechanical system based 
on a {\it finite} number of degrees of freedom \cite{GoldstoneHoppe}.
In a suitable limit to the infinite-dimensional group we thus recover the
supermembrane. This observation enables one to
regularize the supermembrane in a
supersymmetric way by considering a limiting procedure based on a
sequence of groups whose limit yields the area-preserving 
diffeomorphisms. For 
membranes of certain topology it is known how to approximate a 
(sub)group of the area-preserving diffeomorphisms as a particular $N\to 
\infty$ limit of SU($N$). To  
be precise, it can be shown that the structure constants of 
SU($N$) tend to those of the invariant subgroup of the 
diffeomorphisms generated by the exact vectors, up to corrections 
of order $1/N^2$. The structure of the corresponding truncations are
shown in Fig. 1 for a spherical and a toroidal membrane.  While some
of the identities \eqn{identities}  
remain valid at finite $N$, others receive corrections of order 
$1/N^2$. Furthermore, the tensors 
$c^A{}_{\!BC}$ and $c^A{}_{\!B\l}$ are intrinsically undefined at 
finite $N$. Therefore, the expression for $X^-$ is ambiguous for 
the matrix model and Lorentz invariance holds only 
in the large-$N$ limit \cite{DWMN,EMM}. 
We should add  that the matrix regularization works also for the 
case of open supermembranes. In that case one deals with certain 
subgroups of SU($N$). We refer to \cite{open2} for 
further details.

\setlength{\unitlength}{0.5mm}
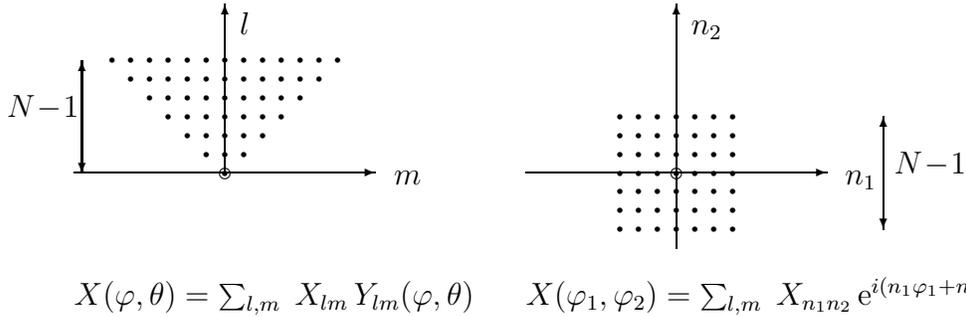
\begin{figure}[t]
\begin{picture}(260,80)(0,0)
\put(20,40){\vector(1,0){80}}
\put(105,37){$m$}
\put(60,40){\vector(0,1){45}}
\put(64,77){$l$}
%
\put(140,40){\vector(1,0){80}}
\put(225,37){$n_1$}
\put(180,20){\vector(0,1){65}}
\put(184,77){$n_2$}
\put(60,40){\circle*{1.5}}
\put(60,40){\circle{3}}
\multiput(55,45)(5,0){3}{\circle*{1.5}}
\multiput(50,50)(5,0){5}{\circle*{1.5}}
\multiput(45,55)(5,0){7}{\circle*{1.5}}
\multiput(40,60)(5,0){9}{\circle*{1.5}}
\multiput(35,65)(5,0){11}{\circle*{1.5}}
\multiput(30,70)(5,0){13}{\circle*{1.5}}
\put(22,40){\vector(0,1){30}}
\put(22,70){\vector(0,-1){30}}
\put(2,55){$N\!-\!1$}
\multiput(165,25)(5,0){7}{\circle*{1.5}}
\multiput(165,30)(5,0){7}{\circle*{1.5}}
\multiput(165,35)(5,0){7}{\circle*{1.5}}
\multiput(165,40)(5,0){7}{\circle*{1.5}}
\multiput(165,45)(5,0){7}{\circle*{1.5}}
\multiput(165,50)(5,0){7}{\circle*{1.5}}
\multiput(165,55)(5,0){7}{\circle*{1.5}}
\put(180,40){\circle{3}}
\put(235,40){\vector(0,1){15}}
\put(235,40){\vector(0,-1){15}}
\put(238,40){$N\!-\!1$}
\put(20,5){$X(\varphi,\theta)= \sum_{l,m} \;X_{lm}\,
Y_{lm}(\varphi,\theta)$}
\put(140,5){$X(\varphi_1,\varphi_2)= \sum_{l,m} \;X_{n_1n_2}\,
{\rm e}^{i (n_1 \varphi_1+n_2 \varphi_2)} $}
%
\end{picture}
\caption{Truncation of spherical harmonics and Fourier modes 
corresponding to an expansion on $S^2$ and $T^2$, respectively. 
The figure shows the case $N=7$. The constant modes associated with
the origin correspond to the U(1) generator while the other 
$N^2-1$ modes are associated with SU($N$).}
\vspace{-4mm}
\end{figure}

The correspondence between the membrane expressions and the matrix
expressions are summarized below:
\be
\begin{array}{rcl}
\int {\rm d}^2\sigma\, \sqrt{w} \; Y_A  = 0 
&\quad\Longleftrightarrow\quad& 
{\rm Tr}\, (T_A) = 0  \\[2mm]
\{Y_A,Y_B\} = f_{AB}{}^{\!C} Y_C  
&\quad\Longleftrightarrow\quad& 
[T_A, T_B]=  f_{AB}{}^{\!C} T_C  \\[2mm]
f_{ABC} = \int {\rm d}^2\sigma\, \sqrt{w} \; Y_A\,\{Y_B,Y_C\} 
&\quad\Longleftrightarrow\quad& 
f_{ABC} ={\rm Tr}\, (T_A [T_B,T_C])  \\[2mm]
\eta_{AB} = \int {\rm d}^2\sigma\, \sqrt{w} \; Y_A Y_B  
&\quad\Longleftrightarrow\quad& 
\eta_{AB}= {\rm Tr}\, (T_AT_B)   \\[2mm]
\int {\rm d}^2\sigma\, \sqrt{w} \; \{Y_A,Y_B\}   = 0 
&\quad\Longleftrightarrow\quad& 
{\rm Tr}\, ([T_A, T_B]) = 0  
\end{array}
\ee

We should stress that the nature of the large-$N$ limit itself is
subtle and is connected to the membrane topology. As long as $N$ is
finite, no distinction can be made with regard to the topology and
clearly the generators of U($N$) as found for different topologies are
related by a simple similarity transformation. In this way one
may establish a mapping between functions on the sphere decomposed into
a finite number of spherical harmonics with $l<N$ and functions on the
torus decomposed into a finite number of Fourier modes belonging to
some fundamental lattice (see Fig.~1). But in
fact there are inequivalent $N\to\infty$ limits. This is in line with
the fact that 
there exists no mapping between differentiable functions on the sphere and
the torus in general, in view of their different topological structure
(cf. the discussion in appendix~B of
\cite{DWMN}). But when taking the trace the precise nature of the 
large-$N$ limit seems less relevant. However, at this point, the
diffeomorphisms associated with the harmonic  
vectors remain problematic; as it turns out they cannot be incorporated for 
finite $N$, at least not at the level of the Lie algebra. This 
was shown in \cite{DWMN}, where it was established that the 
finite-$N$ approximation to the structure constants \tmath{f_{\l 
BC}}  violates the Jacobi identities for a toroidal membrane. 
Therefore it seems impossible to present a matrix
model regularization of the supermembrane with winding 
contributions. There exists a standard prescription for dealing 
with matrix models with winding \cite{Tdual}, however, which 
is therefore conceptually different. The  
consequences of this difference are not well understood. The 
prescription amounts to adopting the gauge group $[{\rm 
U}(N)]^M$, for winding in one dimension, which in the limit $M\to 
\infty$ leads to supersymmetric Yang-Mills
theories in $1+1$ dimensions \cite{Tdual}. Hence, in this way it 
is possible to extract extra dimensions from a suitably chosen 
infinite-dimensional gauge group. This approach can obviously
be generalized to a hypertorus.

\section{Supersymmetric matrix models and their energy spectrum}       
The models that one obtains by a truncation of the gauge group of 
area-preserving diffeomorphisms to a
finite group belong to a class of models proposed long ago as
extended models of supersymmetric quantum mechanics with more
than four supersymmetries \cite{CH}. These theories can also be
obtained from a supersymmetric gauge theory in the zero-volume limit. 
They are based on the Hamiltonian,
\begin{equation}\label{hamiltonian}
H = \frac{1}{g}{\rm Tr}\Big[ \ft{1}{2}{\bf P}^2 - \ft{1}{4}[X^a,X^b]^2
+\ft12 g\,\theta^{\rm T} \gamma_a [ X^a, \theta ]\,\Big] \, ,
\end{equation}
and depend on a number of $d$-dimensional coordinates ${\bf 
X}=(X^1, \ldots,X^d)$, corresponding momenta ${\bf P}$, as well as real 
spinorial anticommuting coordinates $\theta_\alpha$, all taking values in 
the matrix representation of some Lie algebra. The phase space is 
restricted to the subspace invariant  
under the corresponding (compact) Lie group and is therefore subject 
to Gauss-type constraints. These constraints coincide with the 
ones discussed in the previous section.         
The spatial dimension $d$ and the 
corresponding spinor dimension are restricted. The models exist 
for $d=2,3,5$, or 9 dimensions; the (real) spinor dimension equals 
$2, 4, 8$, or 16, respectively. Naturally this is also the number of 
independent supercharges. 

Just as for the supermembrane we restrict ourselves to 
the highest-dimensional case. In that case the model contains 16 
supercharges, denoted by $Q^+$. However, additional charges can 
be obtained when  
the gauge group has abelian factors by including the zero modes 
of the fermion field belonging to the abelian supermultiplet (the 
supercharge of the abelian supermultiplet is already contained in 
the 16 supercharges, in order that one obtains the total Hamiltonian 
from the anticommutator of these supercharges). The extra charges will be
denoted by $Q^-$. Assuming  
one abelian component associated with the matrix trace, we thus 
have 32 supercharges defined by   
\be
Q^+ = {\rm Tr}\Big[(P^a \gamma_a +\ft12 i [X^a,
X^b]\gamma_{ab})\,\theta\,\Big]\,,  \qquad
Q^- = {g}\; {\rm Tr} \left[\, \theta \,\right] \,.
\label{s-charges}
\ee
The $Q^+$ generate the familiar supersymmetry algebra (in the 
group-invariant subspace),
\be
\{Q^+_\alpha,Q^+_\beta\}\approx  H\, \delta_{\alpha\beta} \,
.\label{susyalg} 
\ee
It is possible, though subtle, to also evaluate the central charges of
the supersymmetry algebra for the matrix models \cite{BSS}, which 
at large $N$ tend to the winding charges exhibited in
\eqn{contsusy}. 

As explained in the previous section the supermembrane in the 
light-cone formulation is described by a quantum-mechanical model 
of the type above with an infinite-dimensional gauge group   
corresponding to the area-preserving diffeomorphisms of the 
membrane spacesheet \cite{DWHN} and a coupling constant $g$ given 
by the total light-cone  
momentum $(P_-)_0$. The fact that finite truncations of the gauge 
group are possible allows one to study supermembranes in 
a convenient regularization. The 
connection with the supermembrane shows that the manifest SO(9) 
symmetry, which from the  
viewpoint of the supermembrane is simply the exact transverse 
rotational invariance of the lightcone formulation, extends to 
the full 11-dimensional Lorentz group in the limit of an appropriate  
infinite-dimensional gauge group \cite{DWMN,EMM}.  
 
Classical zero-energy configurations require all commutators
to vanish,
\be
[X^a,X^b]= 0\,. \label{supervalley}
\ee
Dividing out the gauge group implies that zero-energy configurations
are parame\-trized by ${\bf R}^{9N}/S_N$. The zero-energy 
valleys characterized by \eqn{supervalley} extend all the way to
infinity where    
they become increasingly narrow. Their existence raises questions 
about the nature of the spectrum of the Hamiltonian 
\eqn{hamiltonian}. In the bosonic versions of these models the 
wave function cannot freely extend to infinity, because at 
large distances it becomes more and more squeezed 
in the valley. By the uncertainty principle, this gives rise to 
kinetic-energy contributions which increase monotonically  
along the valley. Another way to understand this effect 
is by noting that oscillations perpendicular to the valleys 
give rise to a zero-point energy, which induces an effective 
potential barrier that confines the wave function. 
This confinement causes the spectrum to be discrete. However, for 
the supersymmetric   
models defined by \eqn{hamiltonian} the situation is different. 
Supersymmetry can cause a cancellation of the transverse 
zero-point energy. Then the wave function is no longer confined, 
indicating that the supersymmetric models have a 
continuous spectrum. The latter was rigourously proven for the 
gauge group SU($N$) \cite{DWLN}.    

Whether or not the Hamiltonian 
\eqn{hamiltonian} allows  normalizable or 
localizable zero-energy states, superimposed on the continuous  
spectrum, is a subtle question. Early discussion on the existence 
of such zero-energy states can be found in \cite{DWHN,DWN}; more 
recent discussions can be found in \cite{FH,mboundstates}. 
According to \cite{mboundstates} such states do indeed exist in 
$d=9$. We should emphasize that there is an
important difference   
between states whose energy is exactly equal to zero and states 
of positive energy. The supersymmetry algebra implies that 
zero-energy states are annihilated by the supercharges. 
Hence, they are supersinglets. The positive-energy states, on the 
other hand, must constitute full supermultiplets. So they are 
multiplets consisting of multiples of $128+128$ bosonic $+$
fermionic states (for $d=9$). However, the presence of the extra
suspersymmetry charge causes a further degeneration by $128+128$
states, so that one obtains zero-energy multiplets of 256 states or
positive-energy multiplets comprising (multiples of)  65536 states.

For the supermembrane, the classical zero-mass configurations 
correspond to zero-area stringlike configurations of arbitrary 
length, characterized by the condition that 
\be
\{X^a(\s), X^b(\s)\} = 0\,. 
\ee
As the supermembrane mass is described by a 
Hamiltonian of the type \eqn{hamiltonian}, the mass spectrum of the 
supermembrane is continuous for the same reasons as given 
above. For a supermembrane moving in a target space with  
compact dimensions, winding may raise the mass of the membrane 
state. This is so because winding in more than one direction 
gives rise to a nonzero central charge in the  
supersymmetry algebra, which sets a lower limit on  
the membrane mass. 
This fact should not be interpreted as an indication that 
the spectrum becomes discrete. The possible 
continuity of the spectrum hinges on the two features mentioned 
above. First the system should possess  
continuous valleys of classically degenerate states. 
Qualitatively one recognizes immediately that this feature is not 
directly affected by winding. A classical membrane with 
winding can still have stringlike configurations of arbitrary length,  
without increasing its area. Hence the classical instability 
persists. The second feature is supersymmetry. 
Without winding it is clear that the valley configurations are 
supersymmetric, so that one concludes that the spectrum is 
continuous. With winding the latter aspect is more 
subtle. However, we note that, when the winding density is
concentrated in one part of the spacesheet, then valleys can
emerge elsewhere
corresponding to stringlike configurations with supersymmetry.
Hence, as a space-sheet local field theory,  
supersymmetry can be broken in one region where the winding is 
concentrated and unbroken in 
another. In the latter  region stringlike configurations can form, 
which, at least semiclassically, will not be suppressed by 
quantum corrections \cite{DWPP1}. 
However, in this case we can only describe the 
generic features of the spectrum. These arguments do not  
preclude the existence of mass gaps. Because massless states
exist for the $d=9$ matrix models, we should expect them to exist for 
the supermembrane. In a flat target space these massless states will
constitute massless supermultiplets in 11 spacetime dimensions
and will presumably coincide with supermultiplet of states of
11-dimensional supergravity.  

The continuous mass spectrum of the supermembrane
forms an obstacle in interpretating the membrane states as elementary 
particles, in analogy to what is done in string theory. 
Instead the continuity of the spectrum should be viewed as a  
result of the fact that  
supermembrane states do not really exist as asymptotic states. As we
discussed already in section~1 the membrane collapses into stringlike
configurations and the resulting states are to be 
interpreted as multi-membrane states which possess a continuous mass
spectrum. Qualitatively, the situation for the matrix models 
\eqn{hamiltonian} based on a finite number of degrees of freedom,
is the same as for the supermembrane. 
Among the zero-energy states there are those where the matrices take a 
block-diagonal form, which can be regarded as a direct product of 
states belonging to lower-rank matrix models \cite{BFSS}. The 
fact that the moduli space of ground states, whose nature is 
protected by supersymmetry at the quantum-mechanical level, is 
isomorphic to ${\bf R}^{9N}/S_N$, is already indicative of a 
corresponding description in terms of an $N$-particle Fock space. 

The finite-$N$ matrix models have an independent interpretation 
in string theory. Strings can end on certain defects by means of  
Dirichlet boundary conditions. These defects are 
called D-branes (for further references, see \cite{Dbranes}). 
They can have a $p$-dimensional spatial extension 
and carry Ramond-Ramond charges \cite{Polchinski}. 
D-Branes play an important role in the nonperturbative 
behaviour of string theory. The models of this section are relevant 
for D0-branes (Dirichlet particles). 
The effective short-distance description for D-branes can be 
derived from simple arguments \cite{boundst}. As the strings must be 
attached to the  $p$-dimensional branes, we are dealing with open 
strings whose endpoints are attached to a  
$p$-dimensional subspace. At short distances, the interactions 
caused by these open strings are determined by the massless 
states of the open string, which constitute the ten-dimensional 
Yang-Mills supermultiplet, propagating in a reduced ($p+1$)-dimensional 
spacetime. Because the endpoints of open strings carry Chan-Paton 
factors the effective  
short-distance behaviour of $N$ D-branes can be described in terms 
of a U($N$) ten-dimensional supersymmetric gauge theory reduced 
to the $(p+1$)-dimensional world volume of the D-brane. The 
U(1) subgroup is associated with the center-of-mass motion of the $N$ 
D-branes.                                                     

In the type-IIA superstring one has Dirichlet 
particles moving in a 9-dimensional space. As the world volume 
of the particles is 
one-dimensional ($p=0$), the short-distance interactions between 
these particle is thus described by the model of section~1 with 
gauge group U($N$) and $d=9$. The continuous spectrum without 
gap is natural here, as it is known that, for static D-branes, the 
Ramond-Ramond repulsion cancels against the gravitational and 
dilaton attraction, a similar phenomenon as for BPS 
monopoles. With this gauge group the  
coordinates can be described in terms of $N\times N$ hermitean 
matrices. The valley configurations correspond to the situation 
where all these matrices can be diagonalized simultanously. The 
eigenvalues then define the positions of $N$  
D-particles in the 9-dimensional space. As soon as one or several 
of these particles approach each other 
then the [U(1)]$^N$ symmetry that is left invariant in the 
valley, will be enhanced to a nonabelian 
subgroup of U($N$). Clearly there are more degrees of freedom 
than those corresponding to the D-particles, which are associated 
with the strings stretching between 
the D-particles. As we alluded to above the model naturally 
incorporates configurations corresponding to widely separated 
clusters of D-particles, each of which can be described by a 
supersymmetric quantum-mechanics model based on the product of a 
number of U($k$) subgroups forming a maximal commuting subgroup 
of U($N$). When all the D-particles move further apart this 
corresponds to configurations deeper and deeper into the 
potential valleys.  
These D-particles thus define an independent perspective on the 
models introduced in this section, which can 
be used to study their dynamics. We refer to \cite{Dpart} for  
work along these lines. 

The study of D-branes was further motivated by a conjecture 
according to which the degrees of freedom of M-theory 
are fully captured by the U$(N)$ super-matrix models in the $N\to 
\infty$ limit \cite{BFSS}. The elusive M-theory is defined as the 
strong-coupling limit of type-IIA   
string theory and is supposed to capture all the relevant degrees 
of freedom of all known string theories, both at the perturbative 
and the nonperturbative level \cite{Townsend,witten3}. In 
this description the various string-string dualities are fully 
incorporated. At large distances M-theory is described by 
11-dimensional supergravity. A direct relation between 
supermembranes and type-IIA string theory was emphasized in 
\cite{Townsend}, based on the relation between extremal
black holes in 10-dimensional supergravity \cite{HorStrom} and 
the Kaluza-Klein states of 11-dimensional supergravity 
in an $S^1$ compactification. In this compactification the 
Kaluza-Klein photon coincides with the Ramond-Ramond vector field of 
type-IIA string theory. Therefore Kaluza-Klein states are 
BPS states whose Ramond-Ramond charge is proportional to their 
mass. Hence they have the same characteristics as the Dirichlet 
particles. From this correspondence with the Kaluza-Klein
spectrum one may infer that the corresponding matrix models must
possess zero-energy bound states, whose existence was indeed
established in \cite{mboundstates}. Furthermore, the effective
interaction between   
infinitely many Dirichlet particles must lead to a theory that is 
identical to that of an elementary supermembrane. There are 
alternative compactifications of  
M-theory which make contact with other string theories. 
Supermembranes have been used to provide evidence for the duality 
of M-theory on $\mbox{\bf R}^{10}\times S_1/\mbox{\bf Z}_2$ and 
10-dimensional $E_8\times E_8$ heterotic strings 
\cite{horvw}. Finally let us mention the so-called double-dimensional 
reduction of membranes, which is a truncation that leads to
fundamental strings \cite{DHIS}. Whether this truncation remains
relevant in the context of the full supermembrane theory is an
open question.

\section{Membranes and matrix models in curved space}
So far we considered  supermembranes moving in a flat 
target superspace. Their description follows from substituting
the flat superspace  expressions \eqn{flatssquantities} into the
supermembrane action \eqn{supermem}. However, these expressions can
also be evaluated for nontrivial backgrounds, such as 
those induced by a nontrivial target-space metric, a target-space 
tensor field and a target-space gravitino field, corresponding to 
the fields of (on-shell) 11-dimensional supergravity. This
background can in principle be cast into superspace form by a
procedure known as `gauge completion'~\cite{gc}. For 11-dimensional 
supergravity, the first steps of this procedure were carried 
out long ago~\cite{CF} and recently the results were determined 
to second order in the fermionic coordinates $\theta$ 
\cite{backgr}. 

To elucidate the generic effects of nontrivial backgrounds for
membrane theories, let us confine ourselves for the moment 
to the purely bosonic theory and present
the light-cone formulation of the membrane in a 
background consisting of the metric $G_{\mu\nu}$ and the tensor 
gauge field $C_{\mu\nu\rho}$. In the subsequent sections we will 
include the fermionic coordinates.
The Lagrangian density for the bosonic membrane follows directly 
{}from \eqn{supermem},  
\begin{equation}
{\cal L} = -\sqrt{-g} - \ft{1}{6}\varepsilon^{ijk} \partial_i X^\mu\,
\partial_j
X^\nu \,\partial_k X^\rho\, C_{\rho\nu\mu} \, ,
\end{equation}
where $g_{ij}= \pa_i X^\m \,\pa_j X^\n \,\eta_{\m\n}$.  
In the light-cone formulation, the coordinates are decomposed in 
the usual fashion as $(X^+,X^-,X^a)$ with  
$a=1\ldots 9$. Furthermore we use the diffeomorphisms in the 
target space to bring the metric in a convenient form~\cite{GS},
\begin{equation}
\label{metricgauge}
G_{--}=G_{a-}=0 \, .
\end{equation}
Just as for a flat target space, we identify the time coordinate
of the target space with the world-volume time, by imposing the
condition  $X^+ = \tau$. Moreover we denote the spacesheet
coordinates of the membrane by $\sigma^r$, $r=1,2$.
Following the same steps as for the membrane in flat 
space~\cite{DWHN}, one arrives at a Hamiltonian formulation of 
the theory in terms of coordinates and momenta. These
phase-space variables are subject to a constraint, which 
takes the same form as for the membrane theory in flat space, namely,  
\begin{equation}
\phi_r = P_a \,\partial_r X^a + P_- \,\partial_r X^- \approx 0\, .
\label{constraintmem}
\end{equation}
Of course, the definition of the momenta in terms of the 
coordinates and their derivatives does involve the 
background fields, but at the end all explicit dependence 
on the background cancels out in the phase-space constraints. 

The total Hamiltonian now follows straightforwardly and is equal to 
\bea
H&\!=\!& \int {\rm d}^2\sigma\, \bigg \{
\frac{G_{+-}}{P_- - C_-}\bigg[\ft{1}{2}\Big(P_a-C_a-\frac{P_--C_-}{G_{+-}}
\, G_{a+}\Big)^2+ \ft{1}{4}
(\varepsilon^{rs}\, \partial_r X^a\, \partial_s X^b )^2\bigg]\nonumber\\
&&\hspace{15mm}  -\frac{P_--C_-}{2\, G_{+-}}\, G_{++}- C_+-C_{+-}
+ c^r \phi_r \bigg \} \,. \label{BGHAM}
\eea
where we have included the Lagrange multipliers $c^r$ coupling to the
constraints \eqn{constraintmem}. Observe that transverse indices
are contracted with the metric $G_{ab}$ or its
inverse. Furthermore we have made use  of the folowing
definitions, 
\bea
C_a &=& - \varepsilon^{rs} \partial_r X^- \partial_s X^b \,C_{-ab} +
\ft{1}{2}
\varepsilon^{rs}\partial_r X^b \partial_s X^c \,C_{abc} \, ,      
\nonumber \\
C_{\pm} &=& \ft{1}{2}\varepsilon^{rs}\partial_r X^a \partial_s X^b
\,C_{\pm ab}\,, \nonumber\\
C_{+-} &=& \varepsilon^{rs}\partial_r X^- \partial_s X^a\,
C_{+-a}\,.
\eea

The gauge choice $X^+=\tau$ still allows for $\tau$-dependent 
reparametrizations of the world-space coordinates
$\sigma^r$, which in turn induce transformations on the Lagrange
multiplier $c^r$ through the Hamilton equations of motion.
In addition there remains the freedom of performing tensor gauge
transformations of the target-space three-form $C_{\mu\nu\rho}$.
In order to rewrite \eqn{BGHAM} in terms of a gauge theory of
area-preserving diffeomorphisms it is desirable to obtain a Hamiltonian
which is polynomial in momenta and coordinates.
For this the dynamics of
$P_--C_-$ needs to become trivial, i.e. $\partial_\tau (P_--C_-)=0$, 
allowing us to set it equal to some space-sheet density
$\sqrt{w(\s)}$. The residual invariance group is then constituted by the
area-preserving diffeomorphisms that leave $\sqrt w$ invariant.
The $\tau$-independence of $P_--C_-$ can be achieved by firstly
assuming that the background fields  
are $X^\pm$-independent. Secondly one uses the tensor
gauge transformations to set $C_{-ab}$ equal to a constant
antisymmetric matrix. One then has
\be
\partial_\tau (P_--C_-)\approx \partial_r\, \Bigl [
-\varepsilon^{rs}\partial_s X^a\, C_{+-a} + (P_--C_-)\, c^r\, 
\Bigr ] \,. 
\label{crdef}
\ee
We now choose a gauge such that the right-hand side of this
equation vanishes. In that case the total Hamiltonian takes the
following form, 
\bea
H&\!=\!& \int {\rm d}^2\sigma\, \bigg \{
\frac{G_{+-}}{P_- - C_-}\bigg[\ft{1}{2}\Big(P_a-C_a-\frac{P_--C_-}{G_{+-}}
\, G_{a+}\Big)^2+ \ft{1}{4}
(\varepsilon^{rs}\, \partial_r X^a\, \partial_s X^b )^2\bigg]\nonumber\\
&&\hspace{15mm}  -\frac{P_--C_-}{2\, G_{+-}}\, G_{++}- C_+\nn\\
&&\hspace{15mm} +{1\over{P_--C_-}} \Big[ \varepsilon^{rs} \pa_r
X^a\pa_s X^b \,P_a\,C_{+-b} + C_-\,C_{+-}\Big]  \bigg \} \,, \label{BGham}
\eea
where $P_--C_-\propto \sqrt w$ and $C_{-ab}$ constant. 

At this point one can impose further gauge choices and set
$G_{+-}=1$ and $C_{+-a}=0$. Taking also $C_{-ab}=0$ the
corresponding Hamiltonian can be cast in Lagrangian form 
in terms of a gauge theory of area-preserving diffeomorphisms
\cite{DWPP3},
\bea
{w}^{-1/2}\, {\cal L} &=& \ft{1}{2} (D_0 X^a)^2 + D_0 X^a \left(
\ft{1}{2} C_{abc} \{ X^b, X^c \} + G_{a+} \right) \nonumber \\
&&- \ft{1}{4}\{ X^a, X^b \}^2 + \ft12{G_{++}} + \ft{1}{2}
C_{+ab} \{ X^a, X^b \} \, , \label{BGlagr}
\eea
where the covariant derivatives were introduced in section~3. For
convenience we have set $(P_-)_0 =1$.  
In the case of compact dimensions, it may not always be possible to
set $C_{+-a}$ and $C_{-ab}$ to zero, although they can be
restricted to constants. One can then follow the same procedure as above. 
As alluded to in the first reference of \cite{newBG}, the
Lagrangian then depends explicitly on $X^-$, a feature that was
already exhibited earlier for the winding membrane
(cf. \eqn{gtlagrangian}. However, in the
case at hand, the constraint makes the 
resulting expression for $X^-$ extremely nontrivial. This is clearly
an issue that deserves more study.  Recently the
antisymmetric constant matrix $C_{-ab}$ was conjectured to 
play a role for the matrix model compactification on a
noncommutative torus \cite{newBG}. It should be interesting to
see what the role is of \eqn{BGham} in this context. 

With a reformulation of the membrane in background  fields as a 
gauge theory of area-preserving diffeomorphisms at one's
disposal, one may consider its regularization through a matrix
model by truncating 
the mode expansion for coordinates and momenta along the lines
explained in section~4. This leads to a replacement of Poisson
brackets by commutators, integrals by traces and products of
commuting fields by symmetrized products of the corresponding 
matrices. At that point the original target-space covariance
is affected, as the matrix reparametrizations in terms of
symmetrized products of matrices do not possess a consistent
multiplication structure; 
this is just one of the underlying difficulties in the
construction of matrix models in curved space \cite{oldBG}. 
Finally, one may now study interactions between membranes by
considering the behaviour of a test membrane in a background field
induced by another membrane \cite{interactions}.  
\section{Supergravity in 11 dimensions}
Before moving to the more general superspace backgrounds 
associated with supergravity in 11 spacetime dimensions, we give a
brief summary of this theory in order to establish 
our conventions. The theory is based on an ``elfbein''
field $e_\m{}^{\!r}$, a Majorana gravitino field $\psi_\m$  
and a 3-rank antisymmetric gauge field $C_{\mu\nu\rho}$.  Its 
Lagrangian\footnote{%
  Gamma matrices satisfy $\{\G_r,\G_s\} = 2\eta_{rs}$, where
  $\eta_{rs}$is the tangent-space metric $\eta_{rs}= {\rm
  diag}(-,+,\cdots,+)$. Gamma matrices with multiple indices denote
  antisymetrized products with unit strength. In particular
  $\G^{ r_1r_2\cdots r_{11}} = {\bf 1}\,\varepsilon^{
  r_1r_2\cdots r_{11}}$. The Dirac conjugate is defined by 
  $\bar\psi = i \psi^\dagger\Gamma^0$ for a generic spinor $\psi$. } %
can be written as follows \cite{CJS},
\bea\label{D11-lagrangian}
{\cal L}&\!\!=\!\!&  -\ft12 e\, 
R(e,\omega)  -2 e\,\bar\psi_\m\G^{\m\n\rho}D_\n[\ft12(\omega
+ \hat \omega)]\psi_\rho -\ft1{96}e\, (F_{\m\n\rho\s})^2 \nonumber\\   
&& - \ft1{2\cdot 12^4}  \, \varepsilon^{\m_1\cdots \m_{11}} 
\,F_{\m_1\m_2\m_3\m_4} \,F_{\m_5\m_6\m_7\m_8} \,
C_{\m_9\m_{10}\m_{11}} \nonumber \\
&&- \ft1{96} e \,\Big(\bar\psi_\l \G^{\m\n\rho\s\l\tau} 
\psi_\tau + 12 \,\bar\psi^\m  \G^{\n\rho} \psi^\s\Big) 
  (F + \hat F)_{\m\n\rho\s} \, , 
\eea
where $e= \det e_\m{}^{r}$, 
$\omega_\m{}^{\!rs}$ denotes the spin connection and
$F_{\m\n\rho\s}$ the field strength of the antisymmetric
tensor. A caret denotes that these quantities have been made
covariant with respect to local supersymmetry. The derivative
$D_\mu(\omega)$ is covariant with respect to local Lorentz
transformations,  
\be
D_\m(\omega)\,\e = \Big(\pa_\m -\ft14\omega_\m{}^{rs}
\G_{rs}\Big) \e\, . 
\ee
The supersymmetry transformations are equal to 
\begin{eqnarray}
\delta e_\mu{}^r &=& 2\,\bar\epsilon\, \Gamma^r \psi_\mu \, ,
\nonumber\\
\delta \psi_\mu  &=& D_\mu(\hat\omega) \epsilon + 
T_\mu{}^{\nu\rho\sigma\kappa} \epsilon \,\hat 
F_{\nu\rho\sigma\kappa} \, , \nonumber\\
\delta C_{\mu\nu\rho} &=& -6 \,\bar\epsilon \,\Gamma_{[\mu\nu}
\psi_{\rho]}\, . 
\label{susycomp}
\end{eqnarray}
with 
\begin{equation}
T_r{}^{\! stuv} = \ft{1}{288}
\left( \Gamma_r{}^{\!stuv} - 8\, \delta^{[s}_r
\Gamma^{tuv]} \right)\, ,  \label{def-T}
\end{equation}
and $\hat F_{\m\n\rho\s}$ is the supercovariant field
strength, 
\be
\hat F_{\mu\nu\rho\sigma} = 4\,\partial_{[\mu} C_{\nu\rho\sigma]} 
+ 12 \,\bar\psi_{[\mu}\Gamma_{\nu\rho} \psi_{\sigma]}\,.
\ee
The supercovariant spin connection $\hat\omega^{rs}_\m$ is the one
that corresponds to a vanishing supercovariant torsion
tensor.  

The Lagrangian \eqn{D11-lagrangian} is derived in the context of the 
so-called ``1.5-order'' formalism, in which the spin connection is 
defined as a dependent field determined by its (algebraic) 
equation of motion, whereas its supersymmetry variation in the 
action is treated as if it were an independent field 
\cite{1.5-order}.  Furthermore we note the presence  
of a Chern-Simons-like term $F\wedge F \wedge C$
in the Lagrangian. Under tensor gauge transformations, 
\be
\d_CC_{\m\n\rho} = 3\,\pa_{[\m}\xi_{\n\rho]}\,,
\ee
the corresponding action is thus only invariant up to surface terms.  

We have the following bosonic field equations and Bianchi identities, 
\bea\label{D11-field-eq}
R_{\m\n} &=& \ft1{144}\, g_{\m\n} \,
F_{\rho\s\l\tau}F^{\rho\s\l\tau} -\ft1{12} 
F_{\m\rho\s\l}\,F_\n{}^{\!\rho\s\l}\,, \nonumber\\ 
\pa_{\m}\Big(e \,  F^{\m\n\rho\s}\Big) &=& \ft1{1152} \, 
\varepsilon^{\n\rho\sigma\m_1\ldots\m_{8}} 
F_{\m_1\m_2\m_3\m_4}\,F_{\m_5\m_6\m_{7}\m_{8}}\,, \nonumber \\
\pa_{[\m}F_{\n\rho\s\l]}&=&0\,,
\eea
which no longer depend explicitly on the antisymmetric gauge field. 
An alternative form of the second equation is \cite{page}
\be\label{D11-field-eq2}
\pa_{[\m_1}H_{\m_2\ldots\m_8]} =0\,,
\ee
where $H_{\m_1\ldots\m_7}$ is the dual field strength,
\be
H_{\m_1\ldots\m_7} = \ft1{7!}e\, 
\varepsilon_{\m_1\ldots\m_{11}} 
F^{\m_8\m_9\m_{10}\m_{11}} 
-\ft1{12}  \, F_{[\m_1\m_2\m_3\m_4}\,C_{\m_5\m_6\m_7]}\,.
\ee
When the third equation of \eqn{D11-field-eq} and 
\eqn{D11-field-eq2} receive contributions from certain source
terms on the right-hand side, then the corresponding charges 
can be associated with the
`flux'-integral of  $H_{\m_1\ldots \m_7}$ 
and $F_{\m_1\m_2\m_3\m_4}$ over the boundary of an 8- or a 5-dimensional 
spatial volume, respectively. This volume is transverse to a  
$p=2$ and $p=5$ brane configuration, and the corresponding
charges are 2- and  5-rank Lorentz tensors. For solutions of 
11-dimensional supergravity that contribute to these charges, see
e.g. \cite{sugramem,sugrafive,solirev,Townsend}. 

It is straightforward to evaluate the supersymmetry algebra on 
these fields. The commutator of two supersymmetry transformations 
yields a general-coordi\-nate transformation, a supersymmetry transformation,
a local Lorentz transformation, and a gauge 
transformation associated with the tensor gauge field,
\be
[\d(\e_1),\d(\e_2) ] = \d_{\rm gct} (\xi^\m) + \d(\e_3)
+ \d_L(\l^{rs}) + \d_C(\xi_{\m\n}) \,.
\ee
The parameters of the transformations on the right-hand side 
are given by
\bea
\xi^\m&\!=\!& 2 \,\bar\e_2\G^\m\e_1 \,,\nonumber\\[1.1mm] 
\e_3&\!=\!& -\xi^\m \psi_\m \,, \nonumber\\[.8mm]
\l^{rs} &\!=\!& -\xi^\m\, \hat \omega_\m{}^{rs} + \ft1{72}\, 
\bar\e_2\Big[ \G^{r s\mu\nu\rho\sigma}\hat F_{\mu\nu\rho\sigma} + 24\,   
\G_{\m\nu}\hat F^{rs\mu\nu}\Big]\e_1\,, \nonumber\\
\xi_{\mu\nu} &\!=\!& - \xi^\rho\, C_{\rho\mu\nu}
-2 \,\bar \e_2 \G_{\mu\nu}\e_1\,. \label{compocomm}
\eea

\section{Superspace in terms of component fields}
After the definition of the component fields and transformation rules 
of supergravity in 11 spacetime dimensions, we briefly
introduce the method for constructing  
superspace backgrounds in terms of these component fields. At the
end of this section we present the superspace 
quantities of interest to second order in the anticommuting
coordinates $\theta$ \cite{backgr}. 
The superspace geometry with coordinates $Z^M=(x^\m,\theta^\a)$ is
encoded in the supervielbein 
$E_M{}^{\!A}$ and a spin-connection field $\Omega_M{}^{\!AB}$. In
what follows we will not pay much 
attention to the spin-connection, which is not an independent
field. Furthermore we have an antisymmetric 
tensor gauge field $B_{MNP}$, subject to tensor gauge transformations, 
\be
\d B_{MNP} = 3\,\partial_{[M} \Xi_{NP]}\, .
\ee 
Unless stated otherwise the derivatives with respect to $\theta$
are always left derivatives. 

Under superspace diffeomorphisms corresponding to $Z^M\to
Z^M+\Xi^M(Z)$, the super-vielbein and tensor gauge field
transform as  
\bea
\delta E_M{}^{\!A} &=& \Xi^N \partial_N E_M{}^{\!A} + \partial_M \Xi^N
E_N{}^{\!A} \,,\nonumber \\ 
\delta B_{MNP} &=& \Xi^Q \partial_Q B_{MNP} 
+ 3\,\partial_{[M} \Xi^Q B_{\vert Q\vert NP]}\,.
\eea
Tangent-frame rotations are $Z$-dependent Lorentz
transformations that act on the vielbein according to
\be
\delta E_M{}^A =  \ft{1}{2}
(\Lambda^{rs}L_{rs})^A{}_{\!B}\, E_M{}^{\!B}\,,
\ee
where the Lorentz generators $L_{rs}$ are defined by
\be
\ft{1}{2}(\Lambda^{rs} L_{rs})^t{}_u = \Lambda^t{}_u \, ,\qquad
\ft{1}{2}(\Lambda^{rs} L_{rs})^a{}_b = \ft{1}{4}\Lambda^{rs}
(\Gamma_{rs})^a{}_b\, . 
\ee

The superspace that we are dealing with is not unrestricted but 
is subject to certain constraints and gauge
conditions. Furthermore, we will not describe an
off-shell situation 
as all superfields will be expressed entirely in terms of 
the three component fields of on-shell 11-dimensional supergravity,
the elfbein $e_\m{}^r$, the antisymmetric tensor gauge field
$C_{\m\n\rho}$ and the gravitino field $\psi_\m$. As a result of
these restrictions the residual symmetry transformations are confined to 
11-dimensional diffeomorphisms with parameters $\xi^\m(x)$, 
local Lorentz transformations with parameters $\l^{rs}(x)$, 
tensor-gauge transformations with parameters $\xi_{\m\n}(x)$ and
local supersymmetry transformations with parameters $\e(x)$. 
To derive how the superfields are
parametrized in terms of the component fields it is
necessary to also determine the form of the superspace
transformation parameters, $\Xi^M$,
$\Lambda^{rs}$ and $\Xi_{MN}$, that generate the supersymmetry
transformations. Here it is important to realize
that we are dealing with a gauge-fixed situation. For that reason
the superspace parameters depend on both the $x$-dependent
component parameters defined above as well as on the component 
fields. 
This has two consequences. First of all, local
supersymmetry transformations reside in the superspace
diffeomorphisms, the Lorentz transformations and the
tensor gauge transformations, as $\Xi^M$,
$\Lambda^{rs}$ and $\Xi_{MN}$ are all expected to contain
$\epsilon$-dependent terms. Thus, when considering 
supersymmetry variations of the various fields, one must in
principle include each of the three possible superspace
transformations. Secondly, when 
considering the supersymmetry algebra, it is crucial to
also take into account the variations of the component fields on
which the parameters $\Xi^M$,
$\Lambda^{rs}$ and $\Xi_{MN}$ will depend. 

The method of
casting component results into superspace has a long history and
is sometimes called `gauge completion'. For results in 4
spacetime dimensions 
we refer the reader to \cite{gc}, while results in 11 dimensions
in low orders of $\theta$ were presented in
\cite{CF,backgr}(see also \cite{BrinkHowe}). Here we will follow
\cite{backgr} where results were obtained to second order in $\theta$.  
There are two, somewhat complimentary, ways to obtain information
on the embedding of component fields in superspace geometry. One
is to consider the algebra of the supersymmetry
transformations as generated by the superspace transformations and
to adjust it to the supersymmetry algebra of the component
fields. This determines the 
superspace transformation parameters. The other is to compare the
transformation rules for the  
superfields with the known transformations of the component
fields. This leads to a parametrization of both the superfields
and the transformation parameters in terms
of the component fields and parameters. The evaluation proceeds
order-by-order in 
the $\theta$-coordinates, but at each level one 
encounters ambiguities which can be fixed by suitable
higher-order coordinate redefinitions and gauge choices. The first
step in this iterative procedure is the  
identification at zeroth-order in $\theta$ of some of the component 
fields and transformation parameters with corresponding
components of the superfield
quantities. The underlying assumption is that this identification
can always be implemented by choosing an appropriate gauge. An 
obvious identification is given by 
\cite{AN,BWZ,gc,CF},  
\be
\begin{array}{rcl}
E_\m{}^r(x,\theta=0) &=& e_\m{}^r(x)\,,  \\
E_\m{}^a(x,\theta=0) &=& \psi_\m{}^a(x)\,,\\
B_{\m\n\rho}(x,\theta=0) &=& C_{\m\n\rho}(x)\,, 
\end{array}
\qquad\qquad 
\begin{array}{rcl}
\Xi^\m(x,\theta=0) &=& \xi^\m(x)\,, \\
\Xi^\a(x,\theta=0) &=& \e^\a(x)\,,  \\
\Lambda^{rs}(x,\theta=0) &=& \lambda^{rs}(x)\,, \\
\Xi_{\m\n}(x,\theta=0) &=& \xi_{\m\n}(x)\,.  
\end{array} 
\label{initial}
\ee
As was explained above, the component supersymmetry transformations
with parameters $\e(x)$ are generated by a linear combination of
a superspace diffeomorphism, a local Lorentz and a tensor gauge
transformation; their corresponding parameters will be 
denoted by $\Xi^M(\e)$, $\Lambda^{rs}(\e)$ and $\Xi_{MN}(\e)$,
respectively. Given the embedding of the component fields into
the superfields, application of these specific superspace
transformations should produce the very same  transformation
rules that were defined directly at the component level in the
previous section. The
structure of the commutator algebra of  
unrestricted infinitesimal superspace transformations is
obvious. Two 
diffeomorphisms yield another diffeomorphism, two Lorentz
transformations yield another Lorentz transformation, according to
the Lorentz group structure, while two tensor transformations
commute. On the other hand, a diffeomorphism
and a local Lorentz transformation yield another Lorentz
transformation, and a diffeomorphism and a tensor gauge
transformation yield another gauge transformation. All other
combinations commute. 

For further details we refer to \cite{backgr} and we proceed 
directly to the results. 
For the supervielbein $E_M{}^{\!A}$ the following expressions were
found,  
\bea
E_\m{}^{\!r} &=& e_\m{}^r + 2\, \bar\theta\,\G^r \psi_\m
\nonumber \\
&& 
+ \bar\theta\,\G^r\Big[ -\ft14
\,\hat\omega_\m{}^{\!st}\G_{st} + T_\m{}^{\!\n\rho\s\lambda}\,\hat 
F_{\n\rho\s\lambda}\Big]\theta 
+  {\cal O}(\theta^3)\,, \nonumber \\
E_\m{}^{\!a} &=& \psi_\m{}^{\!a} - \ft14\,
\hat\o_\m^{rs}\,(\G_{rs} \theta)^a + 
(T_\m{}^{\!\n\rho\s\lambda}\theta)^a\, \hat F_{\n\rho\s\lambda} +
{\cal O}(\theta^2)\,, \nonumber \\
E_\a{}^{\!r} &=& -(\bar\theta\,\G^r)_\a +  {\cal O}(\theta^3)\,,
\nonumber \\  
E_\a{}^{\!a} &=& \d^a_\a + M_\a{}^{\!a} + {\cal O}(\theta^3)\,,
\label{svielbein}
\eea
where $M_\a{}^{\!a}$ characterizes the $\hat
F\theta^2$-contributions, which have not been evaluated. Observe that
$E_\m{}^{\!a}$ was determined only up to 
terms of order $\theta^2$. 
The result for the tensor field $B_{MNP}$ reads as follows,
\bea
B_{\m\n\rho} &=& C_{\m\n\rho} -6\, \bar \theta
\G_{[\m\n}\psi_{\rho]}  \nonumber\\
&& -3\,\bar\theta\,\G_{[\m\n}\Big[-\ft14\hat\omega_{\rho]}{}^{\!rs}\,\G_{rs}
+ T_{\rho]}{}^{\!\s\lambda\kappa\tau}\,\hat
F_{\s\lambda\kappa\tau}\Big]\theta  - 12\,
\bar\theta\,\G_{\s[\m} \psi_{\n}\;\bar\theta \,\G^\s \psi_{\rho]}
  +  {\cal O}(\theta^3) \,,\nonumber\\
B_{\m\n\a} &=& (\bar \theta \,\G_{\m\n})_\a - \ft83 \bar\theta\,\G^\rho
\psi_{[\m}\;(\bar\theta\,\G_{\n]\rho})_\a +\ft43
(\bar\theta\,\G^\rho)_\a \;\bar\theta\,\G_{\rho[\m}\psi_{\n]}
+  {\cal O}(\theta^3) \,, \nonumber\\ 
B_{\m\a\b} &=& (\bar\theta\,\G_{\m\n})_{(\a}\,
(\bar\theta\,\G^\n)_{\b)} 
+  {\cal
O}(\theta^3) \, ,  \nonumber \\
B_{\a\b\gamma} &=& (\bar\theta\Gamma_{\m\n})_{(\a}\, 
(\bar\theta\Gamma^\m)_{\b}\, 
(\bar\theta\Gamma^\n)_{\g)}
+  {\cal O}(\theta^3) \,. 
\label{stensor}
\eea
For completeness we included the $\theta^3$-term in
$B_{\a\b\gamma}$ which were already known from the flat-superspace
results \eqn{flatssquantities}. 

Then we turn to some of the transformation parameters. 
The supersymmetry transformations consistent with the fields
specified above, are generated by superspace diffeomorphisms,
local Lorentz transformations and tensor gauge
transformations. The corresponding parameters are as follows. The
superspace diffeomorphisms are expressed by  
\bea
\Xi^\m(\e) &=& \bar\theta\, \G^\m \e -\bar \theta\,
\G^\n\e\;\bar\theta\,\G^\m\psi_\n
+ {\cal O}(\theta^3)\,, \nonumber\\ 
\Xi^\a(\e) &=& \e^\a -\bar\theta \,\G^\m\e\,\psi_\m^\a \nonumber \\
&&+ \bar \theta\,\G^\n\e\; \bar \theta\,
\G^\mu\psi_\n\,\psi_\m{}^{\!\a} +\ft14   \bar\theta\,\G^\n\e \,
\hat\omega^{rs}_\n\,(\G_{rs}\theta)^\a  + \e^\b \,N_\b{}^{\!\a} 
+  {\cal O}(\theta^3)\,, \label{sdiffs}
\eea 
where $N_\b{}^{\!\a}$ encodes unknown terms proportional to $\hat F
\theta^2$. The Lorentz transformation is given by
\be
\Lambda^{rs}(\e) =  \bar\e\,\G^\m\theta\,\hat\omega_\m^{rs} +
\ft1{144} \bar\theta(\G^{rs\m\n\rho\s} \hat F_{\m\n\rho\s} +24 \,\G_{\m\n}
\hat F^{rs\m\n})\e + {\cal O}(\theta^2)\,. \label{slorentz}
\ee
The tensor gauge transformations are parametrized by 
\bea
\Xi_{\m\n}(\e)&=& \bar \e(C_{\m\n\rho}\,\G^\rho + \G_{\m\n}
)\theta 
+\bar\theta\,\G^\rho\e\;\bar\theta(C_{\m\n\s}\,\G^\s +  
\G_{\m\n})\psi_\rho 
+ \ft43 \bar\theta\,\G^\rho\psi_{[\m}\;\bar\theta\,\G_{\n]\rho}\e
\nonumber \\  
&& +\ft43
\bar\theta\,\G^\rho\e \;\bar\theta\,\G_{\rho[\m}\psi_{\n]}
+  {\cal O}(\theta^3)\,, \nonumber \\
\Xi_{\m\a}(\e) &=& \ft16 \bar\theta\,\G^\n \e \,
(\bar\theta\,\G_{\m\n})_\a+ 
\ft16(\theta\,\G^\n)_\a\, \bar\theta\,\G_{\m\n}\e +  {\cal
O}(\theta^3)\,, \nonumber \\ 
\Xi_{\a\b}(\e) &=& {\cal O}(\theta^3) \,.
\label{sgauge}
\eea
Finally local Lorentz transformations are generated by a 
superspace local Lorentz transformation combined with a 
diffeomorphism. The corresponding expressions read 
\bea
\Lambda^{rs}(\lambda)&=&\lambda^{rs}\,, \nonumber\\
\Xi^\a(\lambda)&=& -\ft14\lambda^{rs} (\G_{rs}\theta)^\a\,. 
\eea

\section{The supermembrane in a nontrivial background}
The initial supermembrane action \eqn{supermem} is manifestly covariant
under independent superspace diffeomorphisms, tangent-space
Lorentz transformations  and tensor gauge
transformations. For the specific superspace fields associated
with 11-dimensional on-shell supergravity that we presented in 
the previous section, this is no longer
true and one has to restrict oneself to the superspace
transformations corresponding to the component supersymmetry,
general-coordinate, local Lorentz and tensor gauge 
transformations. When writing \eqn{supermem} in components,
utilizing the expressions found in the previous sections, one
thus obtains an action that is covariant under the restricted
superspace 
diffeomorphisms \eqn{sdiffs} acting on the superspace
coordinates $Z^M=(X^\m,\theta^\a)$ (including the spacetime
arguments of the background fields) combined with usual
transformations on the component fields (we return to this point
shortly). Note that the result 
does not constitute an invariance. Rather it implies that 
actions corresponding to two different 
sets of background fields that are equivalent by a component
gauge transformation, are the same modulo a reparametrization of
the supermembrane embedding coordinates. In order to be precise let us
briefly turn to an example and consider the action of a particle
moving in a curved spacetime background with metric $g_{\m\n}$, 
\be 
S[X^\m, g_{\m\n}(X)]  = - m \int {\rm d}t\; \sqrt{- g_{\m\n}(X(t))\,\dot
X^\m(t)\,\dot X^\n(t)}\,. \label{particle}
\ee
This action, which is obviously invariant under world-line
diffeomorphisms, satisfies $S[X^{\prime\,\m}, 
g^\prime_{\m\n}(X^\prime)]=S[X^\m,g_{\m\n}(X)]$, where
$X^{\prime\,\m}$ and $X^\m$ are related by a target-space general
coordinate transformation which also governs the relation between
$g^\prime_{\m\n}$ and $g_{\m\n}$. Of course, when
considering a background that is invariant under (a subset of 
the) general coordinate transformations (so that $g=g^\prime$), then the
action will be invariant under the corresponding change of the
coordinates. This is the situation that we will address in the next
section, where we take a specific background metric with certain
isometries. In that context the relevant target space for 
\eqn{particle} is an anti-de-Sitter ($AdS_d$) space, which has
isometries that constitute the group SO$(d-1,2)$, where 
$d$ is the spacetime dimension. Then \eqn{particle}
describes a one-dimensional field theory with an SO$(d-1,2)$
invariance group. In the particular case of $d=2$ this invariance can be
re-interpreted as a conformal invariance for a supersymmetric quantum
mechanical system.\footnote{%
  This situation arises generically  for any $p$-brane moving in a
  target space that is locally the product of $AdS_{p+2}$ and some
  compact space. The conformal interpretation was emphasized in
  \cite{KvPTK}}     

Using the previous results one may now write down the complete
action of the supermembrane coupled to background fields up to order
$\theta^2$. Direct substitution leads to the following result for
the supervielbein pull-back,
\bea
\Pi_i^r&=&\partial_iX^\mu\, \Bigl (e_\m{}^r + 2\,
\bar\theta\,\G^r \psi_\m 
-\ft14 \bar\theta\,\G^{rst}\theta\,\hat\omega_{\m\,\!st} + \bar
\theta\,\G^r T_\m{}^{\!\n\rho\s\lambda}\theta \,\hat 
F_{\n\rho\s\lambda} \Big) \nonumber \\
&&
+ \bar\theta\Gamma^r\partial_i\theta + {\cal
O}(\theta^3)\,,\nonumber\\
\Pi_i^a&=& \pa_iX^\m \Big(\psi_\m{}^{\!a} - \ft14\,
\hat\o_\m^{rs}\,(\G_{rs} \theta)^a + 
(T_\m{}^{\!\n\rho\s\lambda}\theta)^a\, \hat
F_{\n\rho\s\lambda}\Big)\nonumber \\
&&+ \pa_i\theta^a  +  {\cal O}(\theta^2)\,.
\label{vielbeinpb}
\eea
Consequently the induced metric is known up to terms of order
$\theta^3$. Furthermore the pull-back of the tensor field equals 
\bea
\lefteqn{-\ft{1}{6}\varepsilon^{ijk}\,
\Pi_i^A\,\Pi_j^B\,\Pi_k^C\, B_{CBA} =  
 -\ft16\vep^{ijk} \pa_iZ^M\,\pa_jZ^N\,\pa_kZ^P\, B_{PNM} =} \nn\\
&& \ft{1}{6}\, dX^{\mu\nu\rho}\, \Big[ 
C_{\m\n\rho} -6\, \bar \theta
\G_{\m\n}\psi_{\rho}
+\ft34\bar\theta\,\G_{rs}\G_{\m\n}\theta\,\hat\omega_{\rho}{}^{\!rs}
\nn\\
&&\hspace{17mm} - 3 \,\bar\theta\,\G_{\m\n}
T_{\rho}{}^{\!\s\lambda\kappa\tau}\theta \,\hat
F_{\s\lambda\kappa\tau} - 12\,
\bar\theta\,\G_{\s\m} \psi_{\n}\;\bar\theta \,\G^\s
\psi_{\rho}\Big]\nonumber\\ 
&&- \varepsilon^{ijk} \,
\bar\theta\,\Gamma_{\mu\nu}\partial_k\theta \Big
[{\textstyle{1\over 2}} 
\,\partial_i X^\mu (\partial_j X^\nu +
\bar\theta\,\Gamma^\nu\partial_j \theta) + {\textstyle{1\over 6}}
\,\bar\theta\,\Gamma^\mu\partial_i\theta\;
\bar\theta\,\Gamma^\nu\partial_j\theta \Big] \nn\\
&& +\ft13 \varepsilon^{ijk} \pa_i X^\m \,\pa_jX^\n \Big[
4\,\bar\theta \,\G_{\rho\m}\pa_k\theta \;
\bar\theta\,\G^\rho\psi_\n - 2 \,\bar \theta\,
\G^\rho\pa_k\theta\; \bar\theta\,\G_{\rho\m}\psi_\n\Big] 
+ {\cal O}(\theta^3)\,, \quad
\label{WZWpb}
\eea
where we have introduced the abbreviation $dX^{\mu\nu\rho}=
\varepsilon^{ijk}\,  
\partial_i X^\mu\,\partial_j X^\nu\, \partial_k X^\rho$ for the
world-volume form. Observe that we included also the 
terms of higher-order $\theta$-terms that were determined in
previous sections and listed in \eqn{stensor}. The first formula
of \eqn{vielbeinpb} and \eqn{WZWpb} now determine the
supermembrane action \eqn{supermem} up to order $\theta^3$.

As an illustration of what we stated at the beginning of this
section, one may consider the effect of the superspace diffeomorphisms
\eqn{sdiffs} on $\Pi^A_i$. We only need the variations to first
order in $\theta$, so that we substitute
$X^\mu\to X^\m+ \bar\theta\G^\m\e$ and $\theta \to \theta + \e -\bar
\theta\,\G^\m\e\;\psi_\m$  into \eqn{vielbeinpb}. For $\Pi^r_i$
this induces a variation which can be rewritten as 
\be
\d\Pi^r_i = \pa_i X^\m \Big[\d e_\m{}^{\!r} +
2\bar\theta\,\G^r\d\psi_\m \Big] - \Lambda^{rs}(\e)
\,\Pi^s_i + {\cal O}(\theta^2)\,.
\ee
The first term on the right-hand side represents the change of
$\Pi^r_i$ under the supersymmetry variations \eqn{susycomp} of
the background fields. The second term represents a Lorentz
transformation whose parameter is given by
\eqn{slorentz}. For the induced metric, given by $g_{ij}=
\Pi_i^r\,\Pi^s_j\,\eta_{rs}$, the Lorentz transformation drops
out, so that the effect of the coordinate change of
$(X^\m,\theta^\a)$ is the same as when performing a supersymmetry
transformation of the background 
fields. This implies that the 
first term in the supermembrane action \eqn{supermem} has indeed
the required transformation behaviour.

A similar result holds for the variation of $\Pi_i^a$ under the
coordinate change as well as for the pull-back of the tensor 
field. Again we refrain from giving further details, but refer
instead to \cite{backgr}. 

While the above results were guaranteed to hold on the basis of
the procedure followed in the previous section, the
$\kappa$-invariance of the action is an independent issue. The
$\kappa$-symmetry transformations are defined in the 
unrestricted superspace and will be given below. In principle, it
should be possible to derive the transformation rules in the
gauge-fixed superspace situation that we are working
with. However, it is not necessary to do so, because we
are only interested in establishing the invariance of the
action. Both the original and the gauge-fixed action should be
$\kappa$-symmetric, so that we can just use the original
superspace diffeomorphisms corresponding to $\kappa$-symmetry and
substitute them in the gauge-fixed 
action. These $\kappa$-transformations take the form of superspace
coordinate changes defined by \cite{BST}
\be
\delta Z^M\, E_M{}^r = 0 \,,\qquad \delta Z^M\, E_M{}^a =
(1-\Gamma)^a{}_b\, \kappa^b \,, \label{kappatransf}
\ee
where $\kappa^a(\zeta)$ is a local fermionic parameter and the
matrix $\Gamma$ is defined by
\be
\Gamma= \frac{\varepsilon^{ijk}}{6\sqrt{-g}}\, \Pi^r_i
\,\Pi^s_j\, \Pi^t_k\, \Gamma_{rst}  \, ,
\label{supergamma}
\ee
with $g=\det g_{ij}$. It satisfies the following properties,
\be
\Gamma^2=1 \,,\qquad \Gamma\,\G_r \Pi^r_i =  \Pi^r_i
\G_r\,\Gamma =  \ft12 {g_{ij}\over
\sqrt{-g}}\,\varepsilon^{jkl}\, \Pi^r_k\,\Pi^s_l\, \G_{rs}
\,. \label{gammaproperties} 
\ee   
Therefore the matrix $(1-\Gamma)$ 
in \eqn{kappatransf} is a projection operator. As a
consequence, this allows one to gauge away half of the $\theta$
degrees of freedom. With these definitions one can prove that the
action is invariant  
under local $\kappa$-symmetry in the appropriate order in 
$\theta$, up to a world-volume surface term which is a 
generalization  of \eqn{kappaboundary}. At this level in $\theta$ there
are as yet no constraints on the background. These constraints will be
required in higher orders of $\theta$ and will take the form of the
supergravity field equations. Again we refer to \cite{backgr} for details.

\section{Near-horizon geometries}
In the previous section we discussed the determination of superspace
quantities,  i.e.  the superspace vielbein and the
tensor gauge field, in terms of the fields of 11-dimensional on-shell
supergravity. The corresponding expressions are obtained by iteration
order-by-order in $\theta$ coordinates, but except for the leading
terms it is hard to proceed with this program. Nevertheless these
results enable one
to write down the 11-dimensional supermembrane action coupled to  
a nontrivial supergravity component-field background to second
order in $\theta$, so that one can start a study of the supermembrane
degrees of freedom in the corresponding  background geometries. In
analogy to the 
bosonic case discussed in section~6, the light-cone supermembrane 
turns out to be equivalent to a gauge theory of area-preserving 
diffeomorphisms coupled to background fields, modulo
corresponding assumptions on the background geometry.
This U($\infty$) gauge theory
may then in turn be regularized by a supersymmetric
U($N$) quantum-mechanical model in curved backgrounds with a certain
degree of supersymmetry. Whether or
not this will shed some light on the problem of formulating
matrix models in curved spacetime is at present still an open
question, as we have already been alluding to in section~6.

However, in specific backgrounds with a certain amount of symmetry, it
is possible to obtain results to all orders in $\theta$. 
Interesting candidates for such backgrounds are the
membrane \cite{sugramem} and the five-brane solution \cite{sugrafive}
of 11-dimensional supergravity,
as well as solutions corresponding to the product of anti-de-Sitter 
spacetimes with compact manifolds \cite{BDPS}. 
Coupling to the latter solutions, which appear near the
horizon of black D-branes \cite{HorStrom}, seem especially
appealing in view of the recent results on a connection between
large-$N$ superconformal field theories and supergravity on a 
product of  AdS space with a compact manifold
\cite{maldacena}. The target-space geometry induced by the $p$-branes
interpolates between $AdS_{p+2}\times B$ near the horizon, where $B$
denotes some compact manifold (usually a sphere), and flat
$(p+1)$-dimensional Minkowski space times a cone with base $B$.

This program has been carried out recently  for the 
type-IIB superstring and the D3-brane
in a IIB-supergravity background of this type 
\cite{MT,KRR,MT2}. In the context of 11-dimensional 
supergravity the $AdS_4\times S^7$ and $AdS_7\times S^4$ 
backgrounds stand out as they leave 32 supersymmetries 
invariant \cite{sevenS,fourS}. These backgrounds  
are associated with the near-horizon geometries corresponding to 
two- and five-brane configurations and thus to possible 
conformal field theories in 3 and 6 spacetime dimensions with 16 
supersymmetries, whose exact nature is not yet completely known. 
In this section we consider the
supermembrane in these two backgrounds \cite{DWPPS}.  As the
corresponding spaces are local products of 
homogeneous spaces, their geometric information can be extracted 
from appropriate coset representatives leading to standard 
invariant one-forms corresponding to the vielbeine and 
spin-connections. The approach of \cite{DWPPS} differs from that
of \cite{DFFFTT}, 
which is also discussed at this conference; in the latter one
constructs the geometric information    
exploiting simultaneously the kappa symmetry of the supermembrane 
action, while in \cite{DWPPS} the geometric information is
determined independently from the supermembrane action. The results
for the geometry coincide with those of \cite{Claus}.

As is well known, the compactifications of the theory to
$AdS_4\times S^7$ and  $AdS_7\times S^4$ are
induced by the antisymmetric 4-rank field strength of
M-theory. 
These two compactifications are thus governed by the 
Freund-Rubin field $f$, defined by (in Pauli-K\"all\'en 
convention, so that we  
can leave the precise signature of the spacetime open),  
\be
F_{\mu\nu\rho\sigma}=6 f \,e\, \varepsilon_{\mu\nu\rho\sigma}\,, 
\label{freundrubin}
\ee
with $e$ the vierbein determinant. 
When $f$ is purely imaginary we are dealing with an $AdS_4\times 
S^7$ background while for real $f$ we have an $AdS_7\times S^4$ 
background. The nonvanishing curvature components corresponding to
the 4- and 7-dimensional subspaces are equal to 
\bea
R_{\mu\nu\rho\sigma}&=&- 4 f^2 ( g_{\mu\rho}\,g_{\nu\sigma}
-g_{\mu\sigma}\,g_{\nu\rho})\,,
\nn\\
R_{\mu'\nu'\rho'\sigma'}&=& f^2 ( g_{\mu'\rho'}\,g_{\nu'\sigma'}
-g_{\mu'\sigma'}\,g_{\nu'\rho'}) \,. \label{curvatures}
\label{sugrasol}
\eea
Here $\m,\n,\rho,\s$ and $\m^\prime,\n^\prime,
\rho^\prime ,\s^\prime$ are 4- and 7-dimensional world indices, 
respectively. We also use $m_{4,7}$ for the inverse radii of the 
two subspaces, defined 
by $\vert f\vert^2={m_7}^2=\ft14{m_4}^2$.
The Killing-spinor equations associated with the 32 
supersymmetries in this background take the form
\bea
\Big(D_\m - f\gamma_\m \gamma_5\otimes {\bf 1} \Big) \epsilon = 
\Big(D_{\m^\prime} +\ft12 f{\bf 1}\otimes\gamma_{\m^\prime}^\prime \Big) 
\epsilon = 0  \,, \label{killing-spinors}
\eea
where we make use of the familiar decomposition of the 
(hermitean) gamma matrices $\gamma_\m$ and
$\gamma^\prime_{\m^\prime}$, appropriate to the 
product space of a 4- and a 7-dimensional subspace (see \cite{DWPPS}).
Here $D_\m$ and $D_{\m^\prime}$ denote the covariant derivatives 
containing the spin-connection fields corresponding to SO(3,1) or 
SO(4) and SO(7) or SO(6,1), respectively. 

The algebra of isometries of the $AdS_4\times S^7$ and 
$AdS_7\times S^4$ backgrounds is given by $osp(8|4;{\bf R})$ and 
$osp(6,2|4)$. Their bosonic subalgebra consists of 
$so(8)\oplus sp(4)\simeq so(8)\oplus so(3,2)$ and $so(6,
2)\oplus usp(4)\simeq so(6,2)\oplus so(5)$, respectively. 
The spinors transform in the $(8,4)$ of this 
algebra. Observe that the spinors transform in a chiral 
representation of $so(8)$ or $so(5)$. 

One may decompose the generators of $osp(8|4)$ or $osp(6,2|4)$ in 
terms of irreducible representations of the bosonic $so(7)\oplus 
so(3,1)$ and $so(6,1)\oplus so(4)$ subalgebras. In that way one
obtains the bosonic (even)  generators $P_r$, $M_{rs}$, which
generate $so(3,2)$ or $so(5)$,  and $P_{r'}$, $M_{r's'}$, which
generate $so(8)$  
or $so(6,2)$. All the bosonic generators are taken antihermitean 
(in the Pauli-K\"all\'en sense). The fermionic 
(odd) generators $Q_{a a'}$ are Majorana spinors, where we denote the
spinorial tangent-space indices by $a,b,\ldots$ and
$a^\prime,b^\prime,\ldots$ for 4- or 7-dimensional indices. 
The commutation relations between even generators are
\be
\begin{array}{rcl}
{[}P_r,P_s{]}&\!=\!&-4f^2 M_{rs}\,,\\[2mm]
{[}P_r,M_{st}{]}&\!=\!&\delta_{rs}\,P_t-\delta_{rt}\,P_s\,,
\\[2mm]
{[}M_{rs},M_{tu}{]}&\!=\!&\delta_{ru}\,M_{st}+\delta_{st}\,
M_{ru}\\[1mm]
&&-\delta_{rt}\,M_{su}-\delta_{su}\,M_{rt}\,,
\end{array}
\begin{array}{rcl}
{[}P_{r'},P_{s'}{]}&\!=\!&f^2 M_{r's'}\,,\\[2mm]
{[}P_{r'},M_{s't'}{]} &\!=\!& \delta_{r's'}\,P_{t'}-\delta_{r't'}\,
P_{s'}\,,\\[2mm]
{[}M_{r's'},M_{t'u'}{]}&\!=\!&\delta_{r'u'}\,M_{s't'}+\delta_{s't'}\,
M_{r'u'}\\[1mm] 
&& -\delta_{r't'}\,M_{s'u'}-\delta_{s'u'}\,M_{r't'}\,.
\end{array} \label{bosonic-comm}
\ee
The odd-even commutators are given by
\be
\begin{array}{rcl}
{[}P_r,Q_{a a'}{]}&\!=\!&- {f} (\gamma_r\gamma_5)_a {}^b \,Q_{b 
a'}\,,\\[2mm]
{[}M_{rs},Q_{a a'}{]}&\!=\!& -\ft12(\gamma_{rs})_a {}^b \,Q_{ba 
'}\,,
\end{array}
\quad
\begin{array}{rcl}
{[}P_{r'},Q_{a a'}{]}&\!=\!& -\ft12 {f}(\gamma'{}_{\!r'})_{a'} {}^{b'} 
\,Q_{a b'}\,,\\[2mm]
{[}M_{r's'},Q_{a a'}{]}&\!=\!& -\ft12 (\gamma'{}_{\!r's'})_{a'} {}^{b'} 
\,Q_{a b'}\,.
\end{array}
\ee
Finally, we have the odd-odd anti-commutators, 
\begin{eqnarray}
\{Q_{a a '},Q_{bb'}\}&=&-(\gamma_5 C)_{a b}
\left(2(\gamma'{}_{\!r'}C')_{a 'b'}\,P^{r'}
- f(\gamma'{}_{\!r's'}C')_{a 'b'}\,M^{r's'}\right) \nonumber\\
&&-C'{}_{\!a' b'}\Bigl(2(\gamma{}_{r}C)_{a b}\,P^{r}
+ 2  f (\gamma{}_{rs}\gamma_5 C)_{a b}\,M^{rs}\Bigr ).
\end{eqnarray}
All other (anti)commutators vanish. The normalizations of the
above algebra were determined by comparison with the 
supersymmetry algebra in the conventions of \cite{backgr} 
in the appropriate backgrounds. 

However, one can return to 11-dimensional notation and drop the
distinction between 4- and 7-dimensional indices so that the equations
obtain a more compact form. In that case the above (anti)commutation
relations that involve the supercharges can be concisely written as,
\begin{eqnarray}
{[}P_{  r},\bar Q{]}&\!=\!& \bar Q \, T_{  r}{}^{\!  s 
  t  u  v} F_{  s  t  u  v} \,,\qquad
{[}M_{  r  s},\bar Q{]}= \ft12 \bar Q\,\Gamma_{  r  s} \,, 
\nonumber\\
\{Q,\bar Q\}&\!=\!&-2 \Gamma_{  r}\,P^{  r}
+ \ft1{144} \Big[ \Gamma^{  r  s  t  u  v  w} 
F_{  t  u  v  w} 
+ 24 \,\Gamma_{  t  u}   F^{  r  s  t  u} \Big] 
M_{  r  s} \,, \label{11-comm}
\end{eqnarray}
where the tensor $T$ is was defined in \eqn{def-T}. Note, however,
that the above 
formulae are only applicable in the background  where the field
strength takes the form given in \eqn{freundrubin}. In what
follows, we will only make use of \eqn{11-comm}. 

\section{Coset-space representatives of $AdS_4\times S^7$ 
and $AdS_7\times S^4$} 
\noindent
Both backgrounds that we consider correspond to homogenous spaces 
and can thus be formulated as coset spaces \cite{cosets}. 
In the case at hand 
these (reductive) coset spaces $G/H$ are $OSp(8|4;{\bf 
R})/SO(7)\times SO(3,1)$ and $OSp(6,2|4)/SO(6,1)\times SO(4)$. 
To each element of the coset $G/H$ one associates an element of 
$G$, which we denote by $L(Z)$. Here $Z^A$ stands for the 
coset-space coordinates $x^{  r}$, $\theta^{  a}$ (or, 
alternatively, $x^r$, $y^{r'}$ and $\theta^{a a '}$). The 
coset representative $L$ transforms from the left under constant 
$G$-transformations corresponding to the isometry group of the 
coset space and from the right under local $H$-transformations: 
$L\to L^\prime = g\,L\,h^{-1}$. 

The vielbein and the torsion-free $H$-connection one-forms, $E$ 
and $\Omega$, are defined through\footnote{%
  A one-form $V$ stands for $V\equiv {\rm d}Z^AV_A$ and an exterior 
  derivative acts according to ${\rm d}V\equiv -{\rm d}Z^B\wedge 
  {\rm d}Z^A \, \partial_AV_B$. Fermionic derivatives are thus 
  always left-derivatives.} 
\be
{\rm d} L + L\,\Omega = L\, E\,,\label{defv}
\ee
where 
\begin{eqnarray}
E= E^{  r}P_{  r} +\bar E Q \,,\qquad 
\Omega= \ft12 \Omega^{  r  s}M_{  r   s}.
\end{eqnarray}
The integrability of \eqn{defv} leads to the Maurer-Cartan 
equations, 
\begin{eqnarray}
\label{maurercartan}
&&{\rm d}\Omega - \Omega \wedge \Omega - \ft12  E^{  r}\wedge 
E^{  s} \, [P_{  r},P_{  s}]
- \ft1{288} \bar E\Big[ \Gamma^{  r  s  t  u  v  w} 
F_{  t  u  v  w} 
+ 24 \,\Gamma_{  t  u}   F^{  r  s  t  u} \Big] 
E\,M_{  r  s} =0  \,, \nn\\
&&{\rm d} E^{  r} -\Omega^{  r  s}\wedge E_{  s} - 
\bar E\,\Gamma^{  r} \wedge E=0 \, ,\nn \\
&&{\rm d} E +  E^{  r}\wedge T_{  r}{}^{  t  u  
v  w}E \,F_{  t   u   v   w} - 
\ft14 \Omega^{  r  s} \wedge \Gamma_{  r  s} E=0 \, ,
\end{eqnarray}
where we suppressed the spinor indices on the anticommuting 
component $E^{  a}$. 
The first equation in a fermion-free background reproduces 
\eqn{curvatures} upon using the commutation relations 
\eqn{bosonic-comm}.   

Now the question is how to determine the vielbeine and
connections to all orders in $\theta$ for the spaces of interest.  
First, observe that the choice of the coset representative amounts to
a gauge choice  
that fixes the parametrization of the coset space. We will 
not insist on an explicit parametrization of the bosonic part of 
the space. It turns out to be advantageous to factorize $L(Z)$ 
into a group element of the bosonic part of $G$ corresponding 
to the bosonic coset space, whose parametrization we leave 
unspecified, and a fermion factor. Hence one may write 
\be
L(Z) =  \ell(x)\, \hat L(\theta)\,, \quad \mbox{ with } \hat 
L(\theta) = \exp [\,\bar \theta Q\,]\,. 
\ee
There exists a convenient trick \cite{MT,KRR,ssp} according to
which one first rescales the odd  
coordinates according to $\theta\rightarrow t\, \theta$, where $t$ is 
an auxiliary parameter that we will put to unity at the end. 
Taking the derivative with respect to $t$ of \eqn{defv} then 
leads to a first-order differential equation for $E$ and $\Omega$ (in 
11-dimensional notation),
\begin{eqnarray}
\dot E - \dot \Omega &=& {\rm d}\bar \theta \,Q  + (E-\Omega) \,
\bar\theta Q - \bar\theta Q \,(E-\Omega) 
\end{eqnarray}
After expanding $E$ and $\Omega$ on the right-hand side in 
terms of the generators and using the (anti)commutation relations 
\eqn{11-comm} we find the coupled first-order linear differential equations, 
\begin{eqnarray}
\dot E^{  a} &=& \Bigl({\rm d}\theta +  E^{  r} \,
T_{  r}{}^{\!  s   t  u  v} \theta \,
F_{  s  t  u  v}  -  \ft14 \Omega^{  r  s}\,
\Gamma_{  r  s}\theta \Bigr)^{  a}\,,\nonumber \\ 
\dot E^{  r} &=& 2\, \bar \theta \,\Gamma^{  r} E\,,\nonumber 
\\
\dot \Omega^{  r  s} &=&  \ft1{72} \bar \theta \Big[ 
\Gamma^{  r  s  t  u  v  w}  
F_{  t  u  v  w} 
+ 24 \,\Gamma_{  t  u}   F^{  r  s  t  u} \Big] 
E\,. \label{diffeq}
\end{eqnarray}
These equations can be solved straightforwardly \cite{KRR} and one finds
\bea
E(x,\theta)&=&\sum_{n=0}^{16}\, \frac{1}{(2n+1)!}\, {\cal M}^{2n} 
\, D\theta\,,  \nn \\ 
E^{  r}(x,\theta)&=&{\rm d}x^{  \mu}\, e_{  \mu} 
{}^{\!  r}  + 2 \sum_{n=0}^{15}\,\frac{1}{(2n+2)!}\, 
\bar\theta\,\Gamma^{  r}\, {\cal M}^{2n}\, D\theta
\\[2.5 mm]
\Omega^{  r  s}(x,\theta)&=&{\rm d}x^{  \mu}\,  
\omega_{  \mu}{}^{\!  r  s}  \nn \\
&& +\ft1{72}  
\sum_{n=0}^{15}\frac{1}{(2n+2)!}\, \bar\theta \,[ 
\Gamma^{  r  s  t  u  v  w}  
F_{  t  u  v  w} 
+ 24 \,\Gamma_{  t  u}   F^{  r  s  t  u} ] 
\, {\cal M}^{2n}\, D\theta    \,, 
\nn
\eea
where the matrix ${\cal M}^2$ equals,
\bea
({\cal M}^2)^{  a}{}_{\!  b}  &=& 2\, (T_{  r}{}^{\!  s 
  t  u  v}\, \theta )^{  a}\, 
F_{s  t  u  v} \, (\bar \theta \,\Gamma^{  
r})_{b} \nn\\
&& - \ft1{288} (\Gamma_{  r  s}\, \theta)^{  
a}\, (\bar\theta\,[\Gamma^{  r  s  t  u  v  w}  
F_{t  u  v  w} 
+ 24 \,\Gamma_{t  u}   F^{  r  s  t  
u}])_{  b}\,.
\eea
and
\be
D\theta^{  a} \equiv \dot E^{  a}\Big\vert_{t=0} =  \Big({\rm d}\theta +  e^{  r} \,
T_{  r}{}^{\!  s   t  u  v} \theta  \, 
F_{  s  t  u  v}  -  \ft14 \omega^{  r  s}\,
\Gamma_{  r  s}\theta\Big)^{  a}\,. 
\ee
It is straightforward to write down the lowest-order terms in these
expansions, 
\begin{eqnarray}
\label{vielbeinexp}
E^{  r} &=& e^{  r} + \bar\theta\Gamma^{  r} {\rm 
d}\theta +  
\bar\theta \Gamma^{  r} ( e^{  m}\,T_{  m}{}^{  s  
t  u  v}  
  F_{  s  t  u  v} - \ft14 \omega^{  s  t} \,
\Gamma_{  s  t})\theta 
+ {\cal O}(\theta^4)\, ,\nn \\
E &=& {\rm d}\theta + 
( e^{  r}\,T_{  r}{}^{  s  t  u  v}
F_{  s  t  u   v} - \ft14 \omega^{  r  s}\, 
\Gamma_{  r  s}) \theta + {\cal O}(\theta^3)   \,,\nn \\
\Omega^{  r  s} &=& \omega^{  r   s}  +\ft1{144}  
 \bar\theta \,[ 
\Gamma^{  r  s  t  u  v  w}  
F_{  t  u  v  w} + 24 \,\Gamma_{  t  u}  
F^{  r  s  t  u} ] \, {\rm d}\theta  
+ {\cal O}(\theta^4) \, , 
\end{eqnarray}
which agree completely with those given in section~8 (and, for the
spin-connection field, in \cite{CF}).  

This information can now be substituted into the first part of the
supermembrane action \eqn{supermem}. 
By similar techniques one can also determine the Wess-Zumino-Witten
part of the action by  first 
considering the most general ansatz for a four-form invariant 
under tangent-space transformations. Using the lowest-order 
expansions of the vielbeine 
(\ref{vielbeinexp}) and comparing with \cite{backgr} shows that 
only two terms can be present. Their relative coefficient is fixed 
by requiring that the four-form is closed, something that can be 
verified by making use of the Maurer-Cartan equations 
(\ref{maurercartan}). The result takes the form 
\begin{equation}
\label{wzwfourform}
F_{(4)} = \frac{1}{4!}\Big[ 
E^{  r}\wedge E^{  s} \wedge E^{  t} \wedge E^{  u} 
F_{  r  s  t  u} 
- 12 \, \bar E \wedge \Gamma_{  r  s} E \wedge E^{  r} 
\wedge E^{  s} \Big]\, .
\end{equation}
To establish this result we also needed the well-known 
quartic-spinor identity in 11 dimensions. The overall factor in 
\eqn{wzwfourform} is fixed by comparing to the normalization of 
the results given in \cite{backgr}.

Because $F_{(4)}$ is closed, it can be written locally  as  
$F_{(4)}= {\rm d} B$. 
The general solution for $B$ can be found by again exploiting the 
one-forms with rescaled $\theta$ coordinates according to $\theta\to t\,
\theta$ and deriving a differential equation. Using the lowest order
result for $B$ this equation can be solved and yields  
\be
B= \ft16\, e^{  r}\wedge e^{  s}\wedge e^{  t} \,C_{  
r   s   t} 
-\int_0^1{\rm d}t\;  \bar\theta \,
\Gamma_{  r  s} E \wedge E^{  r}\wedge E^{  s}\, ,
\ee
where the vielbein components contain the rescaled $\theta$'s. 
This answer immediately reproduces the flat-space result upon
substitution of $F_{r stu}= \omega^{rs}=0$. 

In order to obtain the supermembrane action one substitutes the above
expressions in the action \eqn{supermem}. The resulting action is then
invariant under local fermionic $\kappa$-transformations 
\cite{BST} as well as under the superspace isometries 
corresponding to $osp(8\vert 4)$ or $osp(6,2\vert 4)$.

We have already emphasized that the choice of the coset representative 
amounts to adopting a certain gauge choice in superspace. The choice
that was made in \cite{DWPPS} connects directly to the generic 
11-dimensional superspace results, written in a Wess-Zumino-type 
gauge, in which there is no distinction between spinorial 
world and tangent-space indices. In specific backgrounds, such as 
the ones discussed here, gauge choices are possible which 
allow further simplifications. For this we refer to \cite{KRR} and
other contributions to this volume.  

The results of this section provide a strong independent check of the
low-order $\theta$ results obtained by gauge completion for general
backgrounds \cite{backgr,CF}. 
A great amount of clarity was gained by expressing our results 
in 11-dimensional language, so that both
the $AdS_4\times S^7$ and the $AdS_7\times S^4$ solution could be
covered in one go. Note that in both these  backgrounds the 
gravitino vanishes. 

We have no reasons to expect that the 
11-dimensional form of our results will coincide with the 
expressions for a generic 11-dimensional superspace (with the 
gravitino set to zero) at arbitrary orders in $\theta$.

\section{Concluding remarks}
In this lecture I discussed supermembranes in a variety of situations.
Closed supermembranes can live in flat spaces, or in superspaces
corresponding to supergravity in 11 spacetime dimensions. When the
target space has compact dimensions there is the possibility of
winding. Furthermore open supermembranes exist, though with rather 
restrictive boundary conditions.  In many cases the
supermembrane theory can be 
regularized, resulting in a super matrix model based on a finite number
of degrees of freedom. These are the very same models that describe
the short-distance dynamics of D0-branes. A fascinating feature that
these models share is that their Hilbert space contains both
single-particle and multi-particle states. For the supermembrane the
same feature is present with respect to states with and without winding.

Yet many questions are
still open, as was already stressed in the introduction. For instance,
the nature of the supermembrane spectrum is hard to understand. One
could be tempted and conjecture
that the supermembrane mass spectrum (in flat space) corresponds
simply to the single- and multiple-particle states of supergravity!
At this moment I have no idea how to test the correctness of such a
conjecture. Another open issue concerns the large-$N$ limit of the
super matrix models.

On the more technical side it is gratifying that explicit
constructions of supermembranes in certain nontrivial backgrounds are
now possible. The complete 
M-theory two-brane action in $AdS_4\times S^7$ and $AdS_7\times S^4$
to all orders in $\theta$ represents a further step
in the program of finding the complete anti-de-Sitter background actions 
for the superstring \cite{MT,KRR} and the M2-, D3- \cite{MT2} and
M5-branes initiated for the bosonic part in \cite{KvPTK}. Furthermore,
by studying the interaction between a test membrane in the background
of an M2- or an M5-brane, one may hope to learn more about the
interactions between branes. Some of these issues have already been
considered recently \cite{interactions}.  

The material of this lecture is by no means complete. For instance,
we did not dicuss matrix strings, nor did we review the matrix-model
calculations pertaining to 
supergraviton scattering. Some of these issues are discussed by other
speakers at this workshop.

\vspace{4mm}

\noindent {\bf Acknowledgements}: Much of the material reported in this
lecture resulted from work done in collaborations with Jens Hoppe,
Martin L\"uscher, Uli Marquard, Hermann Nicolai, Kasper Peeters, Jan
Plefka and Alex Sevrin during the past ten years. I thank
them for pleasant and stimulating interactions. I also thank Tom
Banks, Jan de Boer, Michael Douglas, Michael Green, Bernard Julia,
Renata Kallosh, Hirosi Ooguri, Matthias Staudacher, Kelly Stelle and
Paul Townsend for valuable comments and discussions during the past
few years.  
%
%

\end{document}